\documentclass[nocopyrightspace,preprint,numbers,10pt]{sigplanconf}

\usepackage[usenames,dvipsnames]{color}
\usepackage{xcolor}%

\usepackage{mdframed}
\usepackage{amsmath}
\usepackage{amssymb}
\usepackage{listings}
\usepackage{textcomp}
\usepackage{mathtools}
\usepackage{pgfplots}
\usepackage{subfigure}
\usepackage{stmaryrd}
\usepackage{mathbbol}
\usepackage{multirow}
\usepackage{url}
\usepackage{enumerate}
\usepackage{tikz}
\usetikzlibrary{patterns}

\usepackage{etoolbox}

\usepackage{microtype} 

\newcommand*{\mylength}{0.22}
\newcommand*{\mylengthg}{0.6}

\usepackage[ruled, vlined,  commentsnumbered,linesnumbered]{algorithm2e}

\SetCommentSty{mycommfont}
\usepackage{balance}
\usepackage[T1]{fontenc}

\usepackage[scaled=0.87]{beramono}

\usepackage[scaled]{helvet}
\usepackage{pdftexcmds}
\usepackage{pgfplots}
\pgfplotsset{
  /pgfplots/ybar legend/.style={
    /pgfplots/legend image code/.code={%
      \draw[##1,/tikz/.cd,bar width=5pt,yshift=-0.2em,bar 
      shift=0pt]
      plot coordinates {(0cm,0.8em)};
    },
  },
}

\newif\ifapps
\appstrue 

\newcommand{\gcc}{GCC\xspace}

\newcommand{\clang}{Clang\xspace}

\newcommand{\cf}{\hbox{\emph{cf.}}\xspace}

\newcommand{\eg}{\hbox{\emph{e.g.}}\xspace}
\newcommand{\ie}{\hbox{\emph{i.e.}}\xspace}

\newcommand{\wrt}{\hbox{\emph{w.r.t.}}\xspace}

\definecolor{darker-light-gray}{gray}{0.8}

\newtheorem{theorem}{Theorem}

\newtheorem{example}{Example}

\newtheorem{definition}{Definition}

\AtBeginEnvironment{theorem}{\small}
\AtBeginEnvironment{example}{\small}
\AtBeginEnvironment{lemma}{\small}
\AtBeginEnvironment{definition}{\small}

\DeclareRobustCommand{\stirling}{\genfrac\{\}{0pt}{}}
\newcommand{\myie}{\emph{i.e.}}
\newcommand{\myetal}{\emph{et al.}}
\newcommand{\myeg}{\emph{e.g.}}

\newcommand{\mywrt}{\emph{w.r.t.}}

\newcommand{\codeIn}[1]{{\small\tt{#1}}}
\newcommand{\sk}{\mathbb{P}}

\newcommand\mydots{\makebox[0.5em][c]{.\hfil.}}

\newcommand\ignore[1]{}

\newcommand{\anony}[1]{}

\makeatletter

\newcommand*{\anonybugid}[1]{{#1\xspace}
  \stringcases
  {%
    {69951}{1}%
    {69801}{2}%
    {70202}{3}%
    {67619}{4}%
    {69740}{5}%
    {70138}{6}%
    {26973}{7}
    {26994}{8}
    {71405}{9}%
    {28045}{10}%
    {68841}{11}%
    {125}{12}
    {121}{13}
    {10015}{14}
    {1637}{15}
  }%
  {[nada]}%
}

\newcommand{\stringcases}[3]{%
  \romannumeral
  \str@case{#1}#2{#1}{#3}\q@stop
}
\newcommand{\str@case}[3]{%
  \ifnum\pdf@strcmp{\unexpanded{#1}}{\unexpanded{#2}}=\z@
  \expandafter\@firstoftwo
  \else
  \expandafter\@secondoftwo
  \fi
  {\str@case@end{#3}}
  {\str@case{#1}}%
}
\newcommand{\str@case@end}{}
\long\def\str@case@end#1#2\q@stop{\z@#1}
\makeatother

\lstdefinelanguage{scala}{
  morekeywords={abstract,case,catch,class,def,%
    do,else,extends,false,final,finally,%
    for,if,implicit,import,match,mixin,%
    new,null,object,override,package,%
    private,protected,requires,return,sealed,%
    super,this,throw,trait,true,try,%
    type,val,var,while,with,yield},
  otherkeywords={=>,<-,<\%,<:,>:,\#,@},
  sensitive=true,
  morecomment=[l]{//},
  morecomment=[n]{/*}{*/},
  morestring=[b]'',
  morestring=[b]',
  morestring=[b]"''
}

\begin{document}
\toappear{}
\title{Skeletal Program Enumeration for Rigorous Compiler Testing}

\authorinfo{Qirun Zhang \and Chengnian Sun \and Zhendong Su}
           {University of California, Davis, United States}
           {\{qrzhang, cnsun, su\}@ucdavis.edu}

\maketitle

\begin{abstract}

A program can be viewed as a syntactic structure $\mathbb{P}$
(syntactic skeleton) parameterized by a collection of identifiers
$V$ (variable names).  This paper introduces the \emph{skeletal program
  enumeration} (SPE) problem: Given a syntactic skeleton
$\mathbb{P}$ and a set of variables $V$, enumerate a set of programs
$\mathcal{P}$ exhibiting all possible variable usage patterns within
$\mathbb{P}$. It proposes an effective realization of SPE for
systematic, rigorous compiler testing by leveraging three important
observations: (1) Programs with different variable usage patterns exhibit
diverse control- and data-dependence, and help
exploit different compiler optimizations;
(2) most real compiler bugs were revealed by small tests (\myie,
small-sized $\mathbb{P}$) --- this ``small-scope'' observation opens
up SPE for practical compiler validation; and (3) SPE is exhaustive \wrt
a given syntactic skeleton and variable set,
offering a level of
guarantee absent from all existing compiler testing techniques.

The key challenge of SPE is how to eliminate the enormous amount of
equivalent programs \mywrt\ $\alpha$-conversion.  Our main technical
contribution is a novel algorithm for computing the canonical (and
smallest) set of all non-$\alpha$-equivalent programs.  To demonstrate
its practical utility, we have applied the SPE technique to test C/C++
compilers using syntactic skeletons derived from their own regression
test-suites. Our evaluation results are extremely encouraging. In less
than six months, our approach has led to 217 confirmed GCC/Clang bug
reports, 119 of which have already been fixed, and the majority are
long latent despite extensive prior testing efforts. Our SPE algorithm
also provides \emph{six orders} of magnitude reduction. Moreover, in
three weeks, our technique has found 29 CompCert crashing bugs and 42
bugs in two Scala optimizing compilers.  These results demonstrate our
SPE technique's generality and further illustrate its effectiveness.


\end{abstract}

\ccsdesc[500]{Software and its engineering~Software testing and debugging}
\ccsdesc[300]{Software and its engineering~Source code generation}
\ccsdesc[300]{Mathematics of computing~Enumeration}

\keywords
Program enumeration, compiler testing

\section{Introduction}\label{sec:intro}

Compilers are among the most fundamental programming tools for
building software. A compiler bug may result in unintended program
executions and lead to catastrophic consequences for safety-critical
applications. It may also hamper developer productivity as it is
difficult to determine whether an execution failure is caused by
defects in the application or the compiler.  In addition, defects in
compilers may silently affect \emph{all} programs that they compile.
Therefore, improving compiler correctness is crucial.

The predominant approach to validating production compilers consists
of various forms of testing. An important, challenging problem in
compiler testing is \emph{input generation}: How to effectively
generate ``good'' test programs?  Intuitively, a good test program is
productive (\myie, it triggers latent compiler defects) and thorough
(\myie, it stress tests the internal passes of a compiler).  Besides
manually created validation suites (\myeg, Plum Hall~\cite{plumc} and
Perennial~\cite{perennialc}), the main techniques for input program
generation can be categorized as \emph{program generation} or
\emph{program mutation}. Program generation constructs fresh test
programs guided by a language's syntax and semantics. For example,
Csmith is the most well-recognized random program generator for
testing C
compilers~\cite{ChenGZWFER13taming,YangCER11finding}. Program
mutation, on the other hand, focuses on systematically transforming
existing programs. Equivalence Modulo Input (EMI) has been the most
representative mutation-based approach by randomly inserting or
deleting 
code~\cite{LeAS14compiler,LeSS15finding,SunLS16finding}. 
Both approaches are
opportunistic because the typical search space is unbounded, and they
tend to favor large and complex programs.

\paragraph{Skeletal Program Enumeration.}
This paper explores a different, much less explored approach of
\emph{skeletal program enumeration} (SPE) for compiler testing.
Rather than randomly generating or mutating large and complex
programs, is it possible to fully exploit small programs to obtain
bounded guarantees \mywrt\ these small programs?  Specifically, we view
every program $P$ as a syntactic skeleton $\mathbb{P}$ with
placeholders (or holes) for a variables set $V$. Given small sets of
$\mathbb{P}$ and $V$, we obtain new programs $\mathcal{P}$ by
exhaustively enumerating \emph{all} variable usage patterns to fill
the holes in $\mathbb{P}$.  This paper demonstrates its strong
practical utility for compiler testing. Three key observations underlie
our SPE realization:

\begin{itemize}
\item \emph{Most compiler bugs can be exploited through small test
  programs.} According to a recent large-scale study on GCC and
  Clang's bug repositories, each reduced test case in the bug reports
  contains fewer than $30$ lines of code on
  average~\cite{cvqz16toward}. Moreover, in our empirical evaluation
  based on the c-torture test-suite from GCC-4.8.5, each function
  contains only \emph{3} variables with \emph{7} use-def sites
  on average.\footnote{GCC's c-torture test-suite consists of (small)
    test programs that broke the compiler in the past.}

\item \emph{Different variable usage patterns trigger various compiler
  optimization passes.} Consider the programs based on different
  variable usage patterns in Figure~\ref{fig:intro}. Note that the
  programs share the same program skeleton. In $P_1$, a compiler may
  issue a warning on the uninitialized variable \codeIn{a}. In $P_2$,
  due to the constant propagation of \codeIn{b = 1}, variable
  \codeIn{a} is folded to 0 on line 3. Therefore, an optimizing
  compiler performs a dead code elimination of the if
  statement. Finally, in $P_3$, variable \codeIn{b} is folded to $1$
  on line 3. An optimizing compiler then performs constant propagation
  for variable \codeIn{a} on lines $2$ and $4$. Section~\ref{sec:moti}
  illustrates SPE for compiler testing via concrete bugs.

\item \emph{Exhaustive enumeration provides relative guarantees.}
  Given a small syntactic skeleton $P$ with $k$ variables, our
  approach produces input programs for compiler testing by enumerating
  all instances of $P$ exhibiting different variable usage patterns.
  For any programming language, it is also possible to enumerate all
  syntactically valid token sequences (\myie, the syntactic skeletons
  $\mathbb{P}$) up to a given bounded length. Our skeletal program
  enumeration establishes the first step toward realizing bounded
  verification of compilers.\footnote{For languages that allow
  undefined behaviors, such as C/C++, we assume reliable oracles
  exist for detecting undefined behaviors (\cf Section~\ref{sec:ubdis}).} 
\end{itemize}

\newsavebox\myboxintroa
\begin{lrbox}{\myboxintroa}
\begin{minipage}{.2\linewidth}%
\lstset{language=C, mathescape, basicstyle=\scriptsize\ttfamily,  keywordstyle=\bfseries,    xleftmargin=0cm,  numberstyle=\tiny,  numbersep=5pt,   prebreak=\raisebox{0ex}[0ex][0ex]{\ensuremath{\hookleftarrow}},  showstringspaces=false,  upquote=true,  tabsize=2 ,numbers = left,}

\renewcommand{\ttdefault}{pcr}
\begin{lstlisting}
int a,b=1;
b = b-a;
if(a)
  a = a-b;
...
\end{lstlisting}
\end{minipage}
\end{lrbox}

\newsavebox\myboxintrob
\begin{lrbox}{\myboxintrob}
\begin{minipage}{.2\linewidth}%
\lstset{language=C, mathescape, basicstyle=\scriptsize\ttfamily,  keywordstyle=\bfseries,    xleftmargin=0cm,  numberstyle=\tiny,  numbersep=5pt,   prebreak=\raisebox{0ex}[0ex][0ex]{\ensuremath{\hookleftarrow}},  showstringspaces=false,  upquote=true,  tabsize=2, numbers = left,}
\renewcommand{\ttdefault}{pcr}
\begin{lstlisting}
int a,b=1;
a = b-b;
if(a)
  a = a-b;
...
\end{lstlisting}
\end{minipage}
\end{lrbox}

\newsavebox\myboxintroc
\begin{lrbox}{\myboxintroc}
\begin{minipage}{.2\linewidth}%
\lstset{language=C, mathescape, basicstyle=\scriptsize\ttfamily,  keywordstyle=\bfseries,    xleftmargin=0cm,  numberstyle=\tiny,  numbersep=5pt,   prebreak=\raisebox{0ex}[0ex][0ex]{\ensuremath{\hookleftarrow}},  showstringspaces=false,  upquote=true,  tabsize=2, numbers = left,}
\renewcommand{\ttdefault}{pcr}
\begin{lstlisting}
int a,b=1;
a = b-b;
if(b)
  a = b-b;
...
\end{lstlisting}
\end{minipage}
\end{lrbox}

\begin{figure}[!t]

 \subfigure[\mbox{Program $P_1$}]{
\usebox\myboxintroa
   \label{fig:pe1}
} \hfill
\subfigure[\mbox{Program $P_2$}]{
\usebox\myboxintrob
\label{fig:pe2}
}\hfill
\subfigure[\mbox{Program $P_3$}]{
\usebox\myboxintroc
\label{fig:pe3}
}
\caption{Illustrative example for skeletal program enumeration, where we assume that the code snippets are parts of a function.}  \label{fig:intro}
\end{figure}

The essence of SPE is, given a skeleton $\mathbb{P}$ and a set of
variables $V$, producing a set of programs $\mathcal{P}$ by
instantiating each placeholder in the skeleton $\mathbb{P}$ with a
concrete variable $v \in V$. Given a set of $k$ variables and a
$\mathbb{P}$ with $n$ placeholders, a na\"{\i}ve approach produces the SPE
set $\mathcal{P}$ with $k^n$ programs. However, most of the programs
in $\mathcal{P}$ are \emph{$\alpha$-equivalent}, \myie, there exists
an $\alpha$-conversion between any two $\alpha$-equivalent
programs. Since $\alpha$-equivalent programs always exploit the same
control- and data-dependence information, it is redundant to enumerate
them for most purposes and especially for compiler
validation. Generating \emph{only} and \emph{all}
non-$\alpha$-equivalent programs makes SPE a unique and challenging
combinatorial enumeration problem. Existing techniques for enumeration
are inefficient to deal with $\alpha$-equivalence in SPE (please refer
to Section~\ref{sec:pediss} for a detailed discussion).

This paper presents the first practical combinatorial approach
for SPE that generates only non-$\alpha$-equivalent programs in
$\mathcal{P}$. To this end, we formulate SPE as a set partition
problem and tackle the unique challenge of dealing with variable
scoping. As an application of our SPE technique, 
we implement and apply it to test the development versions of GCC and
Clang/LLVM, two popular open-source C/C++ compilers. In less than six
months, we have found and reported 217 bugs,
most of which
are long latent (\eg, more than two thirds of the GCC bugs
affect at least three recent stable releases). About half the bugs
concern C++, an extremely complex language, making our work the first
successful exhaustive technique for testing compilers' C++ support. To further demonstrate its efficiency, we have applied the SPE technique to test CompCert~\cite{Leroy06formal} and two Scala compilers~\cite{dotty11,scala11}. Our three-week testing efforts also yield promising results.

Furthermore, to quantify the effectiveness of our enumeration scheme,
we also apply both our approach and the na\"{\i}ve approach to GCC-4.8.5's
test-suite.  In particular, we use the enumerated programs to test the
stable releases of GCC-4.8.5 and Clang-3.6. Besides finding 11 bugs in
both compilers, more importantly, our approach achieves
six orders of size reduction over the na\"{\i}ve enumeration
approach. Approximately, our approach can process all programs in less than
one month, while the na\"{\i}ve approach would need more than 40K years to
process the same set of test programs.

\paragraph{Contributions.}
Our main contributions follow: 
\begin{itemize}
\item We formulate the problem of skeletal program enumeration
  to aid compiler testing. Unlike existing approaches based on random
  program generation or mutation, our approach exhaustively considers
  all variable usage patterns for small programs;
\item We propose an efficient combinatorial approach to program
  enumeration. In our empirical evaluation, our algorithm reduces the
  search space by six orders of magnitude over na\"{\i}ve enumeration when
  processing compiler test-suites; and 
\item We apply our SPE technique to test GCC, Clang/LLVM, CompCert and two
  Scala compilers. In less than six months, we have found and reported
  217 bugs in GCC and Clang. In about three weeks, we have also found
  29 CompCert crashing bugs, and 42 bugs in the production Scala
  compiler~\cite{scala11} and the Dotty~\cite{dotty11} research
  compiler.  These bugs have been actively addressed by
  developers. For instance, as of November 2016, among our reported GCC bugs, 68\% have
  already been fixed, 66\% are long latent and 10\% are
  release-blocking. 25 CompCert bugs have been fixed and all 27 Dotty bugs have
  been confirmed.
\end{itemize}

\paragraph{Generality.}
Beyond compiler testing, skeletal program enumeration suggests a
general strategy for approaching various enumeration problems. Indeed,
rather than enumerating suitable structures from scratch \wrt\ syntax or
semantics, it can be more profitable to enumerate \wrt\ skeletons
derived from existing structures, which are arguably more interesting
and lead to a more feasible process. Algorithmically, our technique of
casting SPE as the set partition problem and how to support variable
scoping may be adapted to enumeration problems where such information
is relevant, such as functional program enumeration, quantified
formula enumeration, and other domain-specific settings.

\paragraph{Paper Organization.}
The rest of the paper is structured as follows. Section~\ref{sec:moti}
motivates our work via concrete examples, and
Section~\ref{sec:formulation} defines the SPE problem and program
$\alpha$-equivalence. We present our combinatorial program enumeration
algorithm in Section~\ref{sec:pee} and experimental results in
Section~\ref{sec:eval}.  Finally, Section~\ref{sec:rw} surveys related
work, and Section~\ref{sec:ff} concludes.

\section{Motivating Examples}\label{sec:moti}

This section motivates our work using two real compiler bugs found via
SPE: a wrong code bug and a crash bug. A \emph{wrong code bug} is a
compiler miscompilation, \myie, the compiler silently produces a wrong
executable, whose behavior is unintended and different from that of
the original source program. A \emph{crash bug} refers to the compiler
crashing when processing an input program. The wrong code bug is an
example latent bug, and the crash bug was classified as
release-blocking by the GCC developers.

\paragraph{Bug \anonybugid{69951}: \gcc Miscompilation.}
%
%
Figure~\ref{fig:motivating-example-gcc-wrong-code} shows a test
program that triggers a miscompilation in a series of \gcc versions,
ranging from \gcc-4.4 to the latest development trunk (revision
233678).  The bug affects as early as revision 104500 from September
2005, and had been in \gcc even before this revision.  For over ten
years, from then to March 2016, when we found and reported this bug, it had slipped
through various compiler testing techniques and thorough in-house
testing.

\begin{figure}[!t]
    \lstset{language=C,  
      basicstyle=\scriptsize\ttfamily,  
      keywordstyle=\bfseries,  
      numbers=left,  
      xleftmargin=0.2cm,  
      numberstyle=\tiny,  
      numbersep=5pt, 
      breaklines=true,
      showstringspaces=false,  
      upquote=true,  
 commentstyle=\scriptsize\ttfamily\itshape,
      tabsize=2}
    \renewcommand{\ttdefault}{pcr}                      
\begin{lstlisting}                                  
int a = 0;
extern int b __attribute__ ((alias (''a''))); (*@  \label{line:motivating:gcc:wrong:attr} @*)

int main ()
{
  int *p = &a, *q = &b;
  *p = 1;
  *q = 2;  (*@  \label{line:motivating:gcc:wrong:last-write} @*)

//return b;    
  return a; // Bug: the program exits with 1  (*@\label{line:motivating:gcc:wrong:buggy-return} @*)
}
\end{lstlisting}
\caption{This test program is miscompiled by multiple \gcc versions
  from \gcc 4.4 to revision 233678.  This bug affected revision
  104500 in September 2005, and had been latent for over ten years
  until we discovered it via SPE.  The program is expected to return
  2, but incorrectly returns 1 instead.
  }
  \label{fig:motivating-example-gcc-wrong-code}
\end{figure}

This program is expected to exit with 2.  The attribute annotation on
line~\ref{line:motivating:gcc:wrong:attr} declares that the variable
\codeIn{b} is an alias of \codeIn{a}.  As pointers \codeIn{p} and
\codeIn{q} point to \codeIn{a} and \codeIn{b} respectively, they
essentially represent the same memory region (\myie, variable
\codeIn{a}).  The last write to \codeIn{a} is 2 through the pointer
\codeIn{q} on line~\ref{line:motivating:gcc:wrong:last-write}, hence
the exit code of this program should be 2. However, the buggy version of \gcc optimizes the code as if \codeIn{p}
and \codeIn{q} were not aliases, and thus the exit code of this
program becomes 1 instead.  The cause of this bug is that \gcc did not
canonicalize two declarations that share the same memory address
(\myie, \codeIn{a} and \codeIn{b} in this example) into a single one,
thus compromising the soundness of its alias analysis.

This test program is enumerated using a skeleton from GCC's own
test-suite by replacing the original variable \codeIn{b} with
\codeIn{a} on line~\ref{line:motivating:gcc:wrong:buggy-return}. The
program in Figure~\ref{fig:motivating-example-gcc-wrong-code} is
simplified for presentation purposes.  The original program is
slightly larger, and a na\"{\i}ve program enumeration approach generates
3,125 programs. In constrast, our
approach only enumerates 52
non-$\alpha$-equivalent programs,
and exposes the bug.


\paragraph{Bug \anonybugid{69801}: \gcc Internal Compiler Crash.}

Figure~\ref{fig:motivating-example-gcc-crash} shows another bug
example found by SPE.  The test program crashes the development trunk
of \gcc at all optimization levels, including -O0.  The reported bug
has been marked as release-blocking by the GCC developers.

\begin{figure}[!t]
    \lstset{language=C,  
        basicstyle=\scriptsize\ttfamily,  
        keywordstyle=\bfseries,  
        numbers=left,  
        xleftmargin=0.2cm,  
        numberstyle=\tiny,  
        numbersep=5pt,
        commentstyle=\scriptsize\ttfamily\itshape,
        breaklines=true,
        showstringspaces=false,  
        upquote=true,  
        tabsize=2}
    \renewcommand{\ttdefault}{pcr}
    \begin{lstlisting}
struct s { char c[1]; };
struct s a, b, c;
int d; int e;

void bar (void) 
{
//e ? (d==0 ? b : c).c : (e==0 ? b : c).c;(*@\label{line:motivating:gcc:crash:original}@*)
  e ? (d==0 ? b : c).c : (d==0 ? b : c).c;(*@\label{line:motivating:gcc:crash:bug}@*)
}
    \end{lstlisting}
  \caption{This test program crashes the development trunk of \gcc
    (revision 233377) at all optimization levels. The bug has been
    marked as release-blocking.}
    \label{fig:motivating-example-gcc-crash}
\end{figure}

The program is quite simple.  Line~\ref{line:motivating:gcc:crash:bug}
tries to access the field \codeIn{c} via nested conditional
expressions.  This line is also the key to trigger the bug in the
GCC's constant folding pass. \gcc crashes when it is checking whether the second operand (\codeIn{d
  == 0 ? b : c}) and the third operand (\codeIn{d == 0 ? b : c}) are
equal in the function \codeIn{operand\_equal\_p}.  This function
recursively checks whether each component of the two operands are the
same.  When it is checking the integer constant 0 of the binary
expression \codeIn{d == 0}, an assertion is violated because
\codeIn{operand\_equal\_p} is instructed to use the addresses of the
integer constants to test the equality, which is undefined.  In the
bug fix, a flag is set to instruct \codeIn{operand\_equal\_p} to check
integer equality via value comparison.

This test program is also enumerated from GCC's
own test-suite.  The difference between them is shown on
line~\ref{line:motivating:gcc:crash:original}.  The test program is
derived by replacing \codeIn{e} with \codeIn{d} in the third operand
of the whole conditional expression.  This replacement makes the
second and third operands identical, triggering the bug in the
function \codeIn{operand\_equal\_p}.

%
%
%
%
%
%

\section{SPE Problem Formulation}\label{sec:formulation}

This section formalizes skeletal program enumeration.


\subsection{Problem Statement}\label{sec:ps}

\newsavebox\myboxwa
\begin{lrbox}{\myboxwa}
\begin{minipage}{.42\linewidth}%
\lstset{basicstyle=\linespread{1.3}\scriptsize, numbers=none,xleftmargin=1em,  
  basewidth=0.25em,mathescape, breaklines=true,
  literate={->}{$\Coloneqq$}{10}
  {a}{$a$}{1}
  {b}{$b$}{1}
  {S}{$S$}{1}
}
\begin{lstlisting}
a -> $x$ | $n$ | $a_1$ $\mathit{op}_a$ $a_2$
b -> $\mathtt{true}$ | $\mathtt{false}$ | $\mathtt{not}$ $b$ | $b_1$ $\mathit{op}_b$ $b_2$ | $a_1$ $\mathit{op}_r$ $a_2$
S -> $x\coloneqq a$ | $S_1 \,; S_2$ | $\mathtt{while}(b)\;\mathtt{do}\;S$ | $\mathtt{if}(b)\;\mathtt{then}\; S_1\;\mathtt{else}\;S_2$
\end{lstlisting}
\end{minipage}
\end{lrbox}
\newsavebox\myboxwb
\begin{lrbox}{\myboxwb}
\begin{minipage}{.52\linewidth}%
\lstset{basicstyle=\linespread{1.3}\scriptsize, numbers=none,
  xleftmargin=1em,mathescape,basewidth=0.25em, breaklines=true,
  literate={->}{$\Coloneqq$}{10}
  {a}{$a$}{1}
  {as}{$\llbracket$ $a$ $\rrbracket$}{6}
  {cs}{$\llbracket$ $b$ $\rrbracket$}{6}
  {ss}{$\llbracket$ $S$ $\rrbracket$}{6}
  {b}{$b$}{1}
  {S}{$S$}{1}
}
\begin{lstlisting}
as -> $\square$ | $n$ | $\llbracket a_1\rrbracket$ $\mathit{op}_a$ $\llbracket a_2\rrbracket$
cs -> $\mathtt{true}$ | $\mathtt{false}$ | $\mathtt{not}$ $\llbracket b\rrbracket$ | $\llbracket b_1\rrbracket$ $\mathit{op}_b$ $\llbracket b_2\rrbracket$ | $\llbracket a_1\rrbracket$ $\mathit{op}_r$ $\llbracket a_2\rrbracket$
ss -> $\square\coloneqq \llbracket a\rrbracket$ | $\llbracket S_1\rrbracket \,; \llbracket S_2\rrbracket$ | $\mathtt{while}(\llbracket b\rrbracket)\;\mathtt{do}\;\llbracket S\rrbracket$ | $\mathtt{if}(\llbracket b\rrbracket)\;\mathtt{then}\; \llbracket S_1\rrbracket\;\mathtt{else}\;\llbracket S_2\rrbracket$
\end{lstlisting}
\end{minipage}
\end{lrbox}

\begin{figure}

 \subfigure[Syntax rules for $P$.]{
\usebox\myboxwa
   \label{fig:while1}
} \hfill
 \subfigure[Syntax rules for transformed $\mathbb{P}$.]{
\usebox\myboxwb
\label{fig:while2}
}
\caption{Hole transformation for the \textsc{While} language.}  \label{fig:while}
\end{figure}

As mentioned in Section~\ref{sec:intro}, a program $P$ comprises of
two parts: a syntactic skeleton with placeholders for variables, and a
set of variables. We define every usage of a variable in program $P$
as a \emph{hole}, and denote it as $\square$. In particular, let us
consider a \textsc{While}-style language shown in
Figure~\ref{fig:while}. Figure~\ref{fig:while1} gives the syntax rules
for the \textsc{While} language which has been widely used in the program
analysis literature~\cite{0098889princi}. In particular, the
nonterminals $S$, $a$ and $b$ denote statements, arithmetic and
Boolean expressions, respectively. The \textsc{While} language plays a
pivotal role in explaining the basic ideas of our work. Note that the
simple \textsc{While} language does not have scope constraints, and
thus every variable is considered global.

To obtain a program with holes, we recursively apply a hole transformation $\llbracket \rrbracket$ to the \textsc{While} grammar. Figure~\ref{fig:while2} gives the transformed grammar. For any \textsc{While} program $P$, we say $\mathbb{P}$ is a \emph{skeleton} of $P$ iff $T_{\mathbb{P}} = \llbracket T_P \rrbracket$ where $T_{\mathbb{P}}$ and $T_P$ are the respective abstract syntax trees of $\sk$ and $P$. Every hole $\square_i$ in $\sk$ is associated with a hole variable set $\mathbb{v}_i$. The set $\mathbb{v}_i$ describes all variables that belong to the lexical scope of $\square_i$. Therefore, replacing all $\square_i$s in $\sk$ with variables $v \in \mathbb{v}_i$ emits a syntactically valid \textsc{While} program $P^{\prime}$. We say $v \in \mathbb{v}_i$ \emph{fills} $\square_i$, and  $P^{\prime}$ \emph{realizes} $\sk$. A skeleton $\mathbb{P}$ with $n$ holes can be represented as a \emph{characteristic vector} $s_\mathbb{P} = \langle\square_1,\square_2,\ldots,\square_n\rangle$. Therefore, a program $P^{\prime}$ that realizes $\mathbb{P}$ can also be represented as a vector $s_{P^{\prime}} =  \langle v_1, v_2,\ldots,v_n\rangle$ such that $v_i \in \mathbb{v}_i$ fills $\square_i$ in $s_\mathbb{P}$ for all $i \in  [1,n]$.

\newsavebox\myboxpea
\begin{lrbox}{\myboxpea}
\begin{minipage}{.2\linewidth}%
\lstset{mathescape, numbers=none, basicstyle=\scriptsize, xleftmargin=0cm}
\begin{lstlisting}
$a \coloneqq 10;$
$b \coloneqq 1;$
$\mathtt{while}(a)\;\mathtt{do}$
 $a \coloneqq a-b;$
\end{lstlisting}
\end{minipage}
\end{lrbox}
\newsavebox\myboxpeb
\begin{lrbox}{\myboxpeb}
\begin{minipage}{.2\linewidth}%
\lstset{mathescape, numbers=none, basicstyle=\scriptsize, xleftmargin=0cm}
\begin{lstlisting}
$\square \coloneqq 10;$
$\square \coloneqq 1;$
$\mathtt{while}(\square)\;\mathtt{do}$
 $\square \coloneqq \square-\square;$
\end{lstlisting}
\end{minipage}
\end{lrbox}
\newsavebox\myboxpec
\begin{lrbox}{\myboxpec}
\begin{minipage}{.2\linewidth}%
\lstset{mathescape, numbers=none, basicstyle=\scriptsize, xleftmargin=0cm}
\begin{lstlisting}
$b \coloneqq 10;$
$a \coloneqq 1;$
$\mathtt{while}(b)\;\mathtt{do}$
 $b \coloneqq b-a;$
\end{lstlisting}
\end{minipage}
\end{lrbox}
\newsavebox\myboxped
\begin{lrbox}{\myboxped}
\begin{minipage}{.2\linewidth}%
\lstset{mathescape, numbers=none, basicstyle=\scriptsize, xleftmargin=0cm}
\begin{lstlisting}
$a \coloneqq 10;$
$b \coloneqq 1;$
$\mathtt{while}(b)\;\mathtt{do}$
 $b \coloneqq a-b;$
\end{lstlisting}
\end{minipage}
\end{lrbox}

\begin{figure}[!t]

 \subfigure[Program $P$]{
\usebox\myboxpea
   \label{fig:pe1}
} \hfill
\subfigure[Skeleton $\mathbb{P}$]{
\usebox\myboxpeb
\label{fig:pe2}
}\hfill
\subfigure[Program $P_1$]{
\usebox\myboxpec
\label{fig:pe3}
}\hfill
\subfigure[Program $P_2$]{
\usebox\myboxped
\label{fig:pe4}
}
\caption{Program enumeration for the \textsc{While} language.}  \label{fig:pe}
\end{figure}

\begin{definition}[Skeletal Program Enumeration]\label{def:pepe}
Given a skeleton $\sk$ and the hole variable sets $\mathbb{v}_i$ for
each $\square_i$, skeletal program enumeration (SPE) exhaustively
computes a set of programs $\mathcal{P}$, such that each $P \in
\mathcal{P}$ realizes $\sk$.
\end{definition}

\begin{example}\label{ex:1}
Consider the example in Figure~\ref{fig:pe}. Figure~\ref{fig:pe1}
shows a \textsc{While} program $P$, and Figure~\ref{fig:pe2} its
skeleton $\mathbb{P}$ with $6$ holes. Since both $a$ and $b$ are
global variables, we have $\mathbb{v}_1 = \mathbb{v}_2 = \cdots
\mathbb{v}_6 = \{a,b\}$. The program $P_1$ in Figure~\ref{fig:pe3}
realizes $\mathbb{P}$ with $s_{P_1} = \langle b, a, b, b, b,
a\rangle$. Moreover, the program $P_2$ in Figure~\ref{fig:pe4}
realizes $\mathbb{P}$ with $s_{P_2} = \langle a, b, b, b, a,
b\rangle$. Therefore, in the program enumeration of this example, we
have $P, P_1, P_2 \in \mathcal{P}$.
\end{example}

For a skeleton $\sk$ with $n$ holes, program enumeration essentially
generates the $n$-ary Cartesian product  over sets
$\mathbb{v}_1, \mathbb{v}_2,\ldots, \mathbb{v}_n$. As a result, the
search space for generating all possible solutions in $\mathcal{P}$ is
$\prod_{i=1}^{n} |\mathbb{v}_{i}|$, which is clearly exponential in
terms of $n$. For instance, the skeleton $\mathbb{P}$ in
Figure~\ref{fig:pe} realizes $2^6 = 64$ programs, \myie,
$|\mathcal{P}| = 64$.

\subsection{Program $\alpha$-Equivalence}\label{sec:ppi}

The na\"{\i}ve approach to SPE produces an overwhelming amount of programs,
where most of the enumerated instances are equivalent
\mywrt\ $\alpha$-conversion. The $\alpha$-equivalent programs always
exhibit the same control- and data-dependence information. In this
paper, we describe a combinatorial approach to exhaustively enumerate
only non-$\alpha$-equivalent programs in
$\mathcal{P}$. Section~\ref{sec:isoprog} formally defines 
$\alpha$-equivalence using \textsc{While} programs, and
Section~\ref{sec:iande} discusses $\alpha$-equivalence of
practical C programs with scope information.

\subsubsection{$\alpha$-Equivalent  Programs} \label{sec:isoprog}

Let us consider two \textsc{While} programs $P$ and $P_1$ in
Figure~\ref{fig:pe}. The characteristic vectors are $s_{P} = \langle
a, b, a, a, a,b\rangle$ and $s_{P_1} = \langle b, a, b, b, b,
a\rangle$, respectively. As mentioned in Example~\ref{ex:1}, both $P$
and $P_1$ belong to the SPE solution $\mathcal{P}$ in
Figure~\ref{fig:pe}. Particularly, we can transform $P$ to $P_1$ by
replacing all occurrences of variables $a$ and $b$ in $P$ with $b$ and
$a$, respectively. The idea behind the transformation is quite similar
to the concept of $\alpha$-conversion in lambda calculus. It is clear
in Figure~\ref{fig:pe} that $P$ exhibits the same control- and
data-dependence information as $P_1$. Consequently, if $P$ is already
enumerated, there is no need to consider $P_1$.

Let $V$ be the set of all variables in a \textsc{While} program $P$
with $n$ holes. Since the \textsc{While} language does not have
lexical scopes, the set $V$ is the same as the hole variable
set $\mathbb{v}_i$ for each hole $\square_i$, \myie, $V =
\mathbb{v}_1 = \cdots = \mathbb{v}_i$, for all $i \in [1,n]$. Let
$\alpha : V \to V$ be a permutation of set $V$. Given $P$, $V$ and
$\alpha$, we define an \emph{$\alpha$-renaming} such that it replaces
each occurrence of variable $v$ in $P$ with $\alpha(v)$ for all $v \in
V$. The $\alpha$-renaming transforms a program $P$ to $P^{\prime}$,
denoted as $P\xrightarrow{\alpha}P^{\prime}$. For example, in
Figure~\ref{fig:pe}, we have $V = \{a,b\}$, $\alpha =
(\begin{smallmatrix} a&b\\ b&a \end{smallmatrix})$, and
$P\xrightarrow{\alpha}P_1$. Formally, we define 
$\alpha$-equivalence below.

\begin{definition}[Program $\alpha$-Equivalence]\label{def:alphaeq}
Two programs $P_1$ and $P_2$ are $\alpha$-equivalent, denoted as $P_1
\cong P_2$, iff:
\begin{enumerate}[(i).]
\item Both $P_1$ and $P_2$ realize the same $\mathbb{P}$; and 
\item There exists an $\alpha$-renaming such that $P_1\xrightarrow{\alpha}P_2$.
\end{enumerate}

\end{definition}

\begin{example}
Consider Figure~\ref{fig:pe} again. $P$ and $P_1$ are
$\alpha$-equivalent. However, $P$ and $P_2$ are
non-$\alpha$-equivalent programs since their characteristic vectors
are $s_{P} = \langle a,b,a,a,a,b \rangle$ and $s_{P_2} = \langle
a,b,b,b,a,b \rangle$, respectively. It is obvious that there exists no
$\alpha$-renaming between them.
\end{example}

For $\alpha$-equivalent programs $P_1$ and $P_2$, the
$\alpha$-renaming maps the output value of any variable $a$ in $P_1$
to the variable $\alpha(a)$ in $P_2$ for any fixed inputs. Therefore,
the $\alpha$-equivalent \textsc{While} programs are semantically
equivalent. As a result, we can safely eliminate those
$\alpha$-equivalent programs in program enumeration, and thus reduce
the solution set.

\subsubsection{$\alpha$-Equivalence with Scope Information}\label{sec:iande}

\newsavebox\myboxpeaa
\begin{lrbox}{\myboxpeaa}
\begin{minipage}{.48\linewidth}%
\lstset{language=C, mathescape, basicstyle=\scriptsize\ttfamily,  keywordstyle=\bfseries,    xleftmargin=0cm,  numberstyle=\tiny,  numbersep=5pt,     showstringspaces=false,  upquote=true,  tabsize=2, numbers=none}
\renewcommand{\ttdefault}{pcr}
\begin{lstlisting}
int main(){
  int a=1, b=0; 
  if(a){
    int c=3, d=5;
    b = c + d;
  }
  printf("%d", a);
  printf("%d", b);
  return 0;
}
\end{lstlisting}
\end{minipage}
\end{lrbox}
\newsavebox\myboxpeba
\begin{lrbox}{\myboxpeba}
\begin{minipage}{.48\linewidth}%
\lstset{language=C, mathescape, basicstyle=\scriptsize\ttfamily,  keywordstyle=\bfseries,    xleftmargin=0cm,  numberstyle=\tiny,  numbersep=5pt,     showstringspaces=false,  upquote=true,  tabsize=2,numbers=none}
\renewcommand{\ttdefault}{pcr}
\begin{lstlisting}
int main(){
  int $\square$=1, $\square$=0; 
  if($\square$){
    int $\square$=3, $\square$=5;
    $\square$ = $\square$ + $\square$;
  }
  printf("%d", $\square$);
  printf("%d", $\square$);
  return 0;
}
\end{lstlisting}
\end{minipage}
\end{lrbox}
\newsavebox\myboxpeca
\begin{lrbox}{\myboxpeca}
\begin{minipage}{.48\linewidth}%
\lstset{language=C, mathescape, basicstyle=\scriptsize\ttfamily,  keywordstyle=\bfseries,    xleftmargin=0cm,  numberstyle=\tiny,  numbersep=5pt,     showstringspaces=false,  upquote=true,  tabsize=2,numbers=none}
\renewcommand{\ttdefault}{pcr}
\begin{lstlisting}
int main(){
  int c=1, b=0; 
  if(c){
    int a=3, d=5;
    b = a + d;
  }
  printf("%d", c);
  printf("%d", b);
  return 0;
}
\end{lstlisting}
\end{minipage}
\end{lrbox}
\newsavebox\myboxpeda
\begin{lrbox}{\myboxpeda}
\begin{minipage}{.48\linewidth}%
\lstset{language=C, mathescape, basicstyle=\scriptsize\ttfamily,  keywordstyle=\bfseries,    xleftmargin=0cm,  numberstyle=\tiny,  numbersep=5pt,     showstringspaces=false,  upquote=true,  tabsize=2,numbers=none}
\renewcommand{\ttdefault}{pcr}
\begin{lstlisting}
int main(){
  int b=1, a=0; 
  if(b){
    int d=3, c=5;
    a = d + c;
  }
  printf("%d", b);
  printf("%d", a);
  return 0;
}
\end{lstlisting}
\end{minipage}
\end{lrbox}
\newsavebox\myboxpecb

\begin{figure}[!t]

\centering
 \subfigure[Program $P$.]{
\usebox\myboxpeaa
   \label{fig:pec1}
}\subfigure[Skeleton $\mathbb{P}$]{
\usebox\myboxpeba
\label{fig:pec2}
}
\subfigure[Program $P_1$]{
\usebox\myboxpeca
\label{fig:pec3}
}\subfigure[Program $P_2$]{
\usebox\myboxpeda
\label{fig:pec4}
}
\caption{$\alpha$-equivalent C programs.}  \label{fig:peonc}
\end{figure}

The \textsc{While} language in Figure~\ref{fig:pe} does not take
lexical scoping into account. The lexical scope information can 
reduce the size of the SPE set $\mathcal{P}$, even for the
na\"{\i}ve approach. In the remaining sections of this paper, we 
discuss using C programs. However, the conceptual idea is
general and can be adapted to any imperative language.

Let us consider the C programs in Figure~\ref{fig:peonc}. Given a
program $P$ in Figure~\ref{fig:pec1}, we can construct a skeleton
$\mathbb{P}$ shown in Figure~\ref{fig:pec2} and a variable set
$\mathbb{v}_i = \{a, b, c, d\}$ for all $i \in [1, 10]$. The
construction treats all variables as if they were global
variables. According to Definition~\ref{def:pepe}, SPE computes
$4^{10} = 1,048,576$ programs. However, the variables $a$ and $b$ in
$P$ are global variables, while the variables $c$ and $d$ belong to
the local scope of the if statement. Therefore, the variable $a$ can
be used to fill any hole that belongs to $c$, but \emph{not vice
  versa}. With the scope information, a na\"{\i}ve approach only needs to
enumerate $2^5\cdot4^5=32,768$ programs in $\mathcal{P}$.

To cope with lexical scopes in C programs, we extend 
$\alpha$-renaming such that it only maps variables of the same
scope. We define the extended renaming map as a \emph{compact}
$\alpha$-renaming. Moreover, when transforming a C program $P$ to
$\mathbb{P}$, we also associate each hole $\square_i$ and its hole
variable set $\mathbb{v}_i$ in $\mathbb{P}$ with the corresponding
scope information in $P$. Therefore, a hole $\square_i$ can only be
filled with the variables available at the current scope. The variable
types can also be handled by extending the compact $\alpha$-renaming
in a similar way. Finally, it is clear that the compact
$\alpha$-renaming still preserves semantic equivalence.

\begin{theorem}
Given a compact $\alpha$-renaming, and two C programs $P_1$ and $P_2$,
$(P_1 \cong P_2) \implies (P_1 \equiv P_2)$.
\end{theorem}

\begin{example}
In Figure~\ref{fig:peonc}, $P$, $P_1$ and $P_2$ are
$\alpha$-equivalent programs.  In particular, we have
$P\xrightarrow{\alpha_1}P_1$ using an $\alpha$-renaming $\alpha_1 =
(\begin{smallmatrix} a&b&c&d\\ c&b&a&d \end{smallmatrix})$, and
$P\xrightarrow{\alpha_2}P_2$ using a compact $\alpha$-renaming
$\alpha_2 = (\begin{smallmatrix}
  a&b&c&d\\ b&a&d&c \end{smallmatrix})$.  They are all semantically
equivalent, generating the same output ``18''. Moreover, for the
compact $\alpha$-renaming, we have $\mathbb{v}_i = \{a, b\}$ and
$\mathbb{v}_j = \{a, b, c, d\}$, where $i\in\{1,2,3,9,10\}$ and $j
\in \{4,5,6,7,8\}$. As metioned above, the SPE \mywrt\ compact $\alpha$-renamings computes 32 times fewer programs.
\end{example}

\section{SPE Algorithm}\label{sec:pee}

This section presents our combinatorial program enumeration
approach. Our approach only enumerates non-$\alpha$-equivalent
programs. Section~\ref{sec:pebi} describes the main idea based on
programs without scope information. Section~\ref{sec:pets} extends the
idea to handle scope information. Section~\ref{sec:pediss}
provides further relevant discussions.

\subsection{Basic Idea}\label{sec:pebi}

In the SPE problem, the inputs are a syntactic skeleton $\mathbb{P}$
and a set of hole variables $\mathbb{v}_i$. Let us revisit the
example in Figure~\ref{fig:pe}. The skeleton $\mathbb{P}$ in
Figure~\ref{fig:pe2} has $6$ holes. Each hole is associated with the
same hole variable set $\mathbb{v}_i = \{a, b\}$ for all $i
\in [1,6]$. Therefore, there are $2^6$ ways to fill in these holes
using a na\"{\i}ve approach.

As discussed in Section~\ref{sec:ppi}, the $\alpha$-equivalent
programs are redundant for SPE. Having a representative program for
all its $\alpha$-equivalent variants helps reduce the size of the SPE
solution $\mathcal{P}$. Therefore, in our approach, we seek to compute
an SPE set $\mathcal{P}^{\prime}$ of all non-$\alpha$-equivalent
programs, \myie, $P_1 \ncong P_2$ for all distinct $P_1, P_2 \in
\mathcal{P}^{\prime}$.  To realize this, we formulate SPE as a set
partition problem. In particular, we view the $n$ holes in
$\mathbb{P}$ as a set $H = \{1,\ldots,n\}$ of $n$ elements. Filling a
hole with a variable $v \in \mathbb{v}_i$ can also be considered as
partitioning an element $h \in H$ into a subset that corresponds to
$v$. For example, the skeleton $\mathbb{P}$ in Figure~\ref{fig:pe2}
with $6$ holes can be represented as set $H=\{1,\ldots, 6\}$. Let
variable $a$ be the first subset and $b$ the second subset to
partition. The characteristic vector $s_{P_1} = \langle b, a, b, b, b,
a\rangle$ of $P_1$ in Figure~\ref{fig:pe3} can be represented as a set
partition $\{\{1,3,4,5\}, \{2,6\}\}$ of set $H_{P_1}$, where the first
subset represents the holes filled with $b$ and the second subset the
holes filled with $a$. Due to the $\alpha$-equivalence property
mentioned in Section~\ref{sec:isoprog}, the variable names are of no
importance. Therefore, the partition $\{\{1,3,4,5\}, \{2,6\}\}$ is
equivalent to $\{\{2,6\},\{1,3,4,5\}\}$. On the other hand, the
partitions are sensitive to the elements in set $H$ such that
partition $\{\{1,3,4,5\}, \{2,6\}\}$ is different from $\{\{2,3,4,5\},
\{1,6\}\}$.

As a result, given a skeleton $\mathbb{P}$ with $n$ elements and a
hole variable set $\mathbb{v}_i$, where $|\mathbb{v}_i| = k$ for all
$i \in [1,n]$, the SPE problem can be reduced to a combinatorial
problem.

 \begin{quote}
\begin{mdframed}
\small
  Enumerate the ways to partition a set of $n$ elements into $k$
  subsets.
\end{mdframed}
 \end{quote}

\begin{example}
Consider the skeleton $\mathbb{P}$ in Figure~\ref{fig:peonc}. The
characteristic set of $P$ is $\langle a,b,a,c,d,b,c,d,a,b\rangle$. The
corresponding set partition is
$\{\{1,3,9\},\{2,6,10\},\{4,7\},\{5,8\}\}$. $P$, $P_1$ and $P_2$ are
$\alpha$-equivalent, therefore, they have the same set partition.
\end{example}

\subsubsection{Number of Partitions}\label{sec:nop}

In combinatorics, the set partition problems are also known as the
\emph{twelvefold way}, since there are twelve ways to classify all
related problems~\cite{opac-b1132643theart}. When the set elements are
labeled and the subsets unlabeled, the number of ways to partition a
set of $n$ elements into $k$ \emph{non-empty} subsets is denoted by
the \emph{Stirling number of the second
  kind}~\cite{opac-b1132643theart}, denoted as $\stirling{n}{k}$ for
$k \leqslant n$. For $k > n$, we let $\stirling{n}{k} =
\stirling{n}{n}$, \myie, we consider at most $n$ partitions. For our
SPE problem, let $S$ denote the number of all partitions, and we have
\begin{equation}
S = \sum\limits_{i=1}^k {n \brace i}
\end{equation}

For fixed value of $k$, one asymptotic estimation of the Stirling
number of the second kind is $\stirling{n}{k}
\sim\frac{k^{n}}{k!}$~\cite[\S 26.8]{Olver:2010:NHMFnist}. Therefore,
we estimate the SPE solution set as follows:

\begin{equation}
S \sim \frac{1^n}{1!} + \frac{2^n}{2!} + \cdots, + \frac{k^n}{k!}
=O(\frac{k n^k}{k!})=O(\frac{n^k}{(k-1)!})
\end{equation}

The overall complexity of our combinatorial approach is still
exponential. However, it reduces the entire solution set by a notable
constant factor of $(k-1)!$. In practice, it improves the feasibility
of skeletal program enumeration.

\subsubsection{Partition Enumeration}

We adopt the standard approach to enumerate all set partitions in
lexicographic order~\cite{opac-b1132643theart,opac-b1095513comb}. The
conventional approach to encode a unique set partition is using a
\emph{restricted growth
  string}~\cite{opac-b1132643theart,opac-b1095513comb}. For a set of
$n$ elements, a restricted growth string $a_1a_2\ldots a_n$ of length
$n$ satisfies the following: 
\[
a_1=0 \text{ and } a_{i+1} \leqslant 1+\text{max}(a_1, \ldots, a_i) \text{ if } i \in [1,n)
\]
The intuitive meaning of a restricted growth string is that, each
element $h$ in $H$ is partitioned to a subset numbered by $a_i$, where
$i$ represents the index of $h$ in $H$. Moreover, suppose that $m$
elements in $H$ have already been partitioned, if the new element
$m+1$ belongs to a new partition, we always assign the smallest
available number to $a_{m+1}$.
\begin{example}

Consider the skeleton $\mathbb{P}$ in Figure~\ref{fig:pe}. The
characteristic set of $P$ is $\langle a,b,a,a,a,b\rangle$. The
corresponding restricted growth string is ``010001''. Since $P_1$ and
$P$ are $\alpha$-equivalent, their strings are the same. For $P_2$, we
have $s_{P_{2}}=\langle a,b,b,b,a,b \rangle$ and the corresponding
string is ``011101''.
\end{example}

\subsection{Taming Scopes}\label{sec:pets}

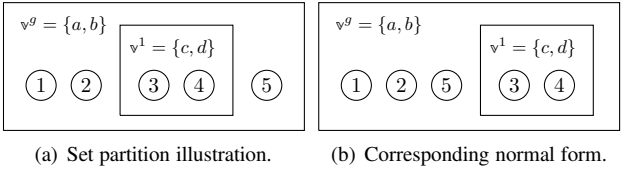
\begin{figure}[!t]
\centering
    \subfigure[Set partition illustration.]{

\begin{tikzpicture}[every path/.style={>=latex}]


  \node [style={ inner sep=0pt, minimum size=0.5cm, scale=0.7}]           (a) at (0.3,0.8)  { $\mathbb{v}^g = \{a, b\}$ };
  \node [style={ inner sep=0pt, minimum size=0.5cm, scale=0.7}]           (a) at (1.75,0.5)  { $\mathbb{v}^1 = \{c, d\}$ };
  \node [style={draw,circle, inner sep=0pt, minimum size=0.5cm, scale=0.8}]           (a) at (0,0)  { $1$ };
  \node [style={draw,circle, inner sep=0pt, minimum size=0.5cm, scale=0.8}]           (a) at (0.6,0)  { $2$ };
  \node [style={draw,circle, inner sep=0pt, minimum size=0.5cm, scale=0.8}]           (a) at (1.5,0)  { $3$ };

  \node [style={draw,circle, inner sep=0pt, minimum size=0.5cm, scale=0.8}]           (a) at (2.1,0)  { $4$ };
  \node [style={draw,circle, inner sep=0pt, minimum size=0.5cm, scale=0.8}]           (a) at (3,0)  { $5$ };

\draw (1.05,-0.4) -- (1.05,0.8) -- (2.55,0.8) -- (2.55,-0.4) -- (1.05,-0.4);

\draw (-0.5,-0.6) -- (-0.5,1.1) -- (3.5,1.1) -- (3.5,-0.6) -- (-0.5,-0.6);

\end{tikzpicture}\label{fig:spi}
}\subfigure[Corresponding normal form.]{

\begin{tikzpicture}[every path/.style={>=latex}]


  \node [style={ inner sep=0pt, minimum size=0.5cm, scale=0.7}]           (a) at (0.3,0.8)  { $\mathbb{v}^g = \{a, b\}$ };
  \node [style={ inner sep=0pt, minimum size=0.5cm, scale=0.7}]           (a) at (2.35,0.5)  { $\mathbb{v}^1 = \{c, d\}$ };
  \node [style={draw,circle, inner sep=0pt, minimum size=0.5cm, scale=0.8}]           (a) at (0,0)  { $1$ };
  \node [style={draw,circle, inner sep=0pt, minimum size=0.5cm, scale=0.8}]           (a) at (0.6,0)  { $2$ };
  \node [style={draw,circle, inner sep=0pt, minimum size=0.5cm, scale=0.8}]           (a) at (1.2,0)  { $5$ };

  \node [style={draw,circle, inner sep=0pt, minimum size=0.5cm, scale=0.8}]           (a) at (2.1,0)  { $3$ };
  \node [style={draw,circle, inner sep=0pt, minimum size=0.5cm, scale=0.8}]           (a) at (2.7,0)  { $4$ };

\draw (1.65,-0.4) -- (1.65,0.8) -- (3.15,0.8) -- (3.15,-0.4) -- (1.65,-0.4);

\draw (-0.5,-0.6) -- (-0.5,1.1) -- (3.5,1.1) -- (3.5,-0.6) -- (-0.5,-0.6);

\end{tikzpicture}\label{fig:nm}
    }
\vspace*{-3pt}
\caption{Set partition illustration and its normal
  form.}\label{fig:spinm}
\end{figure}

The most significant challenge of skeletal program enumeration is to
handle variable scopes. Taking the scope information into
consideration, each hole in the syntactic skeleton $\mathbb{P}$ can be
filled with different sets of variables. As a result, computing the
non-$\alpha$-equivalent programs becomes more difficult. The
corresponding set partition problem of skeletal program enumeration
with scope information is unique and has not been studied in the
literature.

Giving a skeleton $\mathbb{P}$ and hole variable sets $\mathbb{v}_i$
with scope information, we depict the set partition problem using a
figure with circle and squares, where each labeled circle denotes the
corresponding $\square_i \in \mathbb{P}$ and the squares represent the
scope information. In particular, according to the compact
$\alpha$-renaming described in Section~\ref{sec:iande}, each hole
(circle) can only be filled with the variables from a valid scope
(square). We use the notations $\mathbb{v}^g$ and $\mathbb{v}^l$ to
represent the sets of global variables and the variables declared in
scope $l$, respectively. Consider the example in Figure~\ref{fig:spi}. We have
$\mathbb{v}^g = \{a, b\}$, and $\mathbb{v}^1 = \{c, d\}$ for the
first local scope. It is clear from the figure that hole $2$ can be
filled with $v \in \mathbb{v}_2 = \mathbb{v}^g$ whereas hole $4$ can
be filled with $v \in \mathbb{v}_4 = \mathbb{v}^g \cup
\mathbb{v}^1$. It is also clear that a na\"{\i}ve approach generates
$2^3\cdot4^2$ solutions.

\subsubsection{Set Partition for Skeletal Program Enumeration}\label{sec:spforpe}

Unlike the standard set partition problem discussed in
Section~\ref{sec:pebi}, the new enumeration problem essentially
considers the set partition of a set $H$ with constraints on each
element $h \in H$. This section formalizes this new
partition problem.

Consider a program skeleton $\mathbb{P}$ with $n$ holes, and $t$
scopes. Each hole $\square_i \in \mathbb{P}$ can be either global or
local.  The global hole $\square_i^g$ can be filled with only global
variables, \myie, $\mathbb{v}_i = \mathbb{v}^g$ whereas the local
hole $\square_i^l$ can be filled with additional local variables
defined in scope $l$, \myie, $\mathbb{v}_i = \mathbb{v}^g \cup
\mathbb{v}^l$ and $l \in [1,t]$. The set partition problem for SPE
can be described as follows:

\begin{quote}
\begin{mdframed}
\small
Given a set $H$ of $n$ elements, and pre-defined sets $\mathbb{v}_i
\subseteq \{1, \ldots, k\}$ for all $i \in [1, n]$. Each element $i
\in H$ can be partitioned to a subset labeled by $v \in \mathbb{v}_i$
and $\mathbb{v}_1 \cup \mathbb{v}_2 \cup \ldots \cup \mathbb{v}_i =
\{1, \ldots,k\}$. Enumerate the ways to partition $H$ into $k$
subsets.
\end{mdframed}
\end{quote}

\subsubsection{Partitions with Scopes}\label{sec:pws}

A straightforward approach to compute partitions with scopes is
computing a local set partition solution $S_l$ for each scope,
respectively. Then, obtain the final solution $S$ by computing the
Cartesian product over all local solutions $S_l$ together with the
solution of global holes $S_g$, \myie, $S = S_g \times S_1 \times
\cdots \times S_t$. However, the set partitions obtained are not the
global solution among all elements. For example, consider the holes
$3$ and $4$ in Figure~\ref{fig:spi}, where we have $\mathbb{v}_3 =
\mathbb{v}_4 = \mathbb{v}^g \cup \mathbb{v}^1$. The local solution
for them contains two partitions: $\{\{3, 4\}\}$ and
$\{\{3\},\{4\}\}$. Since the number of holes is smaller than the
number of hole variables, we pick two variables and let $\mathbb{v}_3
= \mathbb{v}_4 = \{b, c\}$. Locally, the variable names are unimportant
in set partition problems. Therefore, for partition
$\{\{3\},\{4\}\}$, filling variables $\square_3 \leftarrow b$ and
$\square_4 \leftarrow c$ is equivalent to $\square_3 \leftarrow c$ and
$\square_4 \leftarrow b$. On the other hand, combining the two
solutions with the remaining holes filled with $\langle a,a, b\rangle$
obtains two solutions $\langle a, a, b, c, b\rangle$ and $\langle a,
a, c, b, b\rangle$. Clearly, they are two unique solutions since they
have different restrict growth strings ``00121'' and ``00122'',
respectively.

To obtain the global solution, the key idea in our partition algorithm
is to choose some local holes by considering all combinations of the
local holes. Then, the chosen ones are promoted to be global. Finally,
we obtain the solution by computing the Cartesian product of the
global holes and the remaining local holes.

\begin{procedure}[!h]
\scriptsize
  \SetKwFunction{curset}{curset}
\SetKwFunction{Up}{Up}
\SetKwFunction{Set}{Set}
\SetKwFunction{PL}{ParititionScope}
\DontPrintSemicolon
\BlankLine

$u \leftarrow |l_i|$ and $v \leftarrow |\mathbb{v}_i|$ \nllabel{proc:0}\\
\ForEach{$k \in [0, u-1]$}{
  $\mathit{result} \leftarrow \textsc{Combinations}(l_i, k)$ \nllabel{proc:1}\\
  \ForEach{variable set $l \in \mathit{result}$}{
  $G  \leftarrow G \cup l$ \nllabel{proc:3}\\
  $\overline{l} \leftarrow l_i \setminus l$ \nllabel{proc:2}\\
  \ForEach{$j \in [1, v]$}{
    $S_{l_{i}} \leftarrow \textsc{Partitions$^{\prime}$}(\overline{l}, j)$ \nllabel{proc:7}\\
    $S_L^{\prime} \leftarrow S_L$\\
    \leIf{$S_L$ is empty}{$S_L \leftarrow S_{l_{i}}$}{$S_L \leftarrow S_L \times S_{l_{i}}$}   \nllabel{proc:9}
    \eIf{$i$ is not the last scope}{
      \PL$(S_L, G, l_{(i+1)})$ \nllabel{proc:11}\\
    }{
      $S_G \leftarrow \textsc{Partitions$^{\prime}$}(G, |\mathbb{v}^g|)$ \nllabel{proc:13}\\
      $S_f \leftarrow S_f \cup \{S_G \times S_L\}$ \nllabel{proc:14}
    }
    $S_L \leftarrow S_L^{\prime}$ \nllabel{proc:15}
    
  }
  
  $G \leftarrow G \setminus l$ \nllabel{proc:16}
}
}

  \caption{PartitionScope ($S_L, G, l_i$).} \label{proc:one}
\end{procedure}

\paragraph{Handling one scope.}
Procedure \codeIn{PartitionScope}$(S_L,G, l_i)$ describes the major
steps for handling scope $i$, where $S_L$ represents the set of all
local solutions, $G$ denotes the set of the global holes and $l_i$ the
set of the local holes of scope $i$. The routine
\textsc{Combinations}$(Q,k)$ returns ${|Q|\choose k}$ different ways
of selecting $k$ elements from the set $Q$. The routine
\textsc{Partitions}$(Q,k)$ partitions the set $Q$ into $k$ subsets in
$\sum\nolimits_{i=1}^k \stirling{|Q|}{k}$ ways.  Moreover, the routine
\textsc{Partitions$^{\prime}$}$(Q,k)$ partitions the set $Q$ into $k$
non-empty subsets in $\stirling{|Q|}{k}$ ways.  \codeIn{PartitionScope} handles a scope $i$ as follows:
\begin{itemize}
\item \emph{Promoting $k$ holes from scope $i$.} Line~\ref{proc:0}
  obtains the cardinalities of sets $l_i$ and $\mathbb{v}_i$. On
  line~\ref{proc:1}, we choose $k$ holes from $l_i$ and promote them
  as global holes $G$ on line~\ref{proc:3}. The set of remaining holes
  is denoted as $\overline{l}$ on line~\ref{proc:2}.

\item \emph{Computing local solution for scope $i$.}
  Lines~\ref{proc:7}-\ref{proc:9} computes the local set partition
  $S_{l_i}$ of scope $i$ and combines it with the current local
  solution $S_L$. If $i$ is not the last scope, it recursively handles
  the next scope $i+1$ on line~\ref{proc:11}.

\item \emph{Obtaining the final solution.} If $i$ is the last scope,
  it computes the solution $S_G$ of global holes $G$
  (line~\ref{proc:13}), combines it with the current local solution
  $S_L$ and appends it to the final solution $S_f$
  (line~\ref{proc:14}). On line~\ref{proc:15} and~\ref{proc:16}, the
  information on $G$ and $S_L$ is restored for subsequent recursive
  calls to Procedure \codeIn{PartitionScope}.
\end{itemize}

\begin{algorithm}[!t]
\scriptsize
\SetKwData{Null}{Null}
\SetKwData{oldIn}{oldIn}
\SetKwData{Up}{up}
\SetKwData{current}{current$_u$}
\SetKwData{previous}{previous$_u$}
\SetKwData{lastc}{last$_u$}
\SetKwData{lastp}{last$_p$}
\SetKwData{outaf}{\textsc{Out}$_{\alpha}$}
\SetKwData{outd}{out$_d$}
\SetKwData{outdb}{out$_{\bar{d}}$}
\SetKwData{rowu}{Row$_u$}

\SetKwFunction{Add}{Add}
\SetKwFunction{Init}{Init}
\SetKwFunction{PF}{ParititionFunc}
\SetKwInOut{Input}{Input}
\SetKwInOut{Output}{Output}
\DontPrintSemicolon

\Input{A program skeleton $\mathbb{P}$ and hole variable set $\mathbb{v}_i$;  }
\Output{a set of programs $\mathcal{P}$.}

\BlankLine
\ForEach{function $f \in \mathbb{P}$ \nllabel{algo:for}}{

Normalize function $f$ \nllabel{algo:2}\\

$S_f^{\prime} \leftarrow \textsc{Partitions}(H_f, |\mathbb{v}_f|)$ \nllabel{algo:3}\\

$S_f \leftarrow \emptyset$ and $S_L \leftarrow \emptyset$ \nllabel{algo:4}\\
\PL$(S_L, G_f, l_1)$ \nllabel{algo:5}\\
$S_f \leftarrow S_f \cup S_f^{\prime}$ \nllabel{algo:6}

  $S \leftarrow S \times S_f$ \nllabel{algo:7}

}
\ForEach{characteristic vector $s \in S$}{
  Generating a program $P$ using $\mathbb{P}$ and $s$

}

\caption{Skeletal program enumeration algorithm.}\label{algo:pe}

\end{algorithm}

\paragraph{Program enumeration algorithm.}
Algorithm~\ref{algo:pe} describes our combinatorial SPE algorithm. For
each function $f$ in skeleton $\mathbb{P}$, we consider its
characteristic vector $s_f = \langle 1,\ldots,n \rangle$. Within a
function $f$, the global variable set $\mathbb{v}_f$ contains the
global variables in $\mathbb{P}$, function parameters and
function-wise variables. Moreover, the set of global holes, denoted as
$H_f$, contains the holes that can be filled with $v \in
\mathbb{v}_f$. For $t$ local scopes, we rearrange the vector to be of
the \emph{normal form} $\langle \square^g,\mydots,\square^g,
\square^1,\mydots ,\square^1,\ldots,\square^t,\mydots,\square^t
\rangle$, \myie, we pull all global holes to the front and arrange
local holes in order. For example,
Figure~\ref{fig:nm} gives the normal form of the holes in
Figure~\ref{fig:spi}. Let $G_f$ and $l_i$ be the sets of the global
holes in $f$ and the local holes in scope $i$,
respectively. Algorithm~\ref{algo:pe} computes the partitions for
function $f$ as follows. It normalizes $f$ (line~\ref{algo:2}) and
computes a partial solution for $f$ (line~\ref{algo:3}) without taking
scopes into consideration. It then computes a solution $S_f$ by
recursively processing each scope on
lines~\ref{algo:4}-\ref{algo:5}. Moreover, the global solution of
function $f$ is obtained by combining both $S_f$ and $S_f^{\prime}$ on
line~\ref{algo:6}. The global solution of $\mathbb{P}$ is obtained by
computing the Cartesian product of each function on
line~\ref{algo:7}. Finally, we enumerate the programs according to the
solutions in $S$.

\begin{example}
Consider the normal form in
Figure~\ref{fig:nm}. Algorithm~\ref{algo:pe} computes the set
partitions as follows.
\emph{Computing $S_f^{\prime}$}: There are
  $\stirling{5}{2}+\stirling{5}{1}=16$ partitions;
\emph{Promoting either $3$ or $4$}: There are
  $\stirling{4}{2}\times\stirling{1}{1}=7$ partitions for each hole;
\emph{Promoting neither $3$ nor $4$}: There are
  $\stirling{3}{2}\times(\stirling{2}{2}+\stirling{2}{1})=6$
  partitions;
\emph{Final solution}: SPE algorithm computes  $(16+2\cdot7+6)=36$ partitions.
However, the na\"{\i}ve approach computes $(2^3\cdot4^2)=128$ partitions.
\end{example}





\subsection{Discussions}\label{sec:pediss}

\paragraph{Granularity of enumeration.}
Algorithm~\ref{algo:pe} obtains the SPE solution $S$ of a skeleton
$\mathbb{P}$ by computing the Cartesian product \mywrt\ each local
solution $S_f$ of function $f$. We say that Algorithm~\ref{algo:pe}
computes the \emph{intra-procedural} enumeration. Since each function
can also be considered as a local scope \mywrt\ a program, the
intra-procedural enumeration approximates the global solution, where
we call the global solution as the \emph{inter-procedural}
enumeration. Algorithm~\ref{algo:pe} can be easily extended to obtain
the inter-procedural enumeration. The key extension is to replace the
{\bf foreach} loop on lines~\ref{algo:for}-\ref{algo:7} with a call to
Procedure \codeIn{PartitionScope}, where $l_1$ represents the first
function scope instead. To handle additional scopes, one can processes
all scopes in a bottom-up fashion \mywrt\ the scope hierarchy. It is a
practical design choice of enabling intra- or inter-procedural
enumerations. The intra-procedural enumeration --- though being an
approximation --- enumerates fewer variants of a single test program
$P$ than the inter-procedural counterpart. Thus, given a fixed
budget on the total number of enumerated variants, the
intra-procedural enumeration is able to process more test programs.
It would be interesting to investigate different
enumeration granularities and find the most cost-effective enumeration
scheme for practical use.

\paragraph{Enumeration vs.\ counting.}
We have discussed the SPE set partition problem, and proposed an enumeration algorithm. An
interesting open problem is to investigate the corresponding counting
counterpart of the enumeration problem in
Section~\ref{sec:spforpe}. Specifically, fixing $i$ and $k$ in an SPE
problem, the counting problem is to determine the number of
non-$\alpha$-equivalent programs for a syntactic skeleton $\mathbb{P}$
with $n$ holes. In Section~\ref{sec:nop}, we discussed the counting
problem of the SPE problem without scope information, based on the
traditional analysis of set partition
problems~\cite{opac-b1095513comb,opac-b1132643theart}. However,
developing an asymptotic estimation of the SPE problem defined
in Section~\ref{sec:spforpe} is nontrivial, as the analytics with the
variable set $\mathbb{v}_i$ constraints becomes more complex. A
promising direction may be counting the enumeration set using the
technique based on e-restricted growth
functions~\cite{MansourN08gray,MansourNV11loop}.

\paragraph{Other enumeration techniques.}
Algorithm~\ref{algo:pe} solves the SPE problem based on the
combinatorial algorithms for generating set partitions and
combinations. In the literature, there has been an extensive body of
work that exhaustively generates input structures for software
testing. This line of work typically specifies the \emph{invariant}
property and enumerates the structures
declaratively~\cite{BoyapatiKM02korat,KhurshidM04testera,SenniF12genera,GaleottiRPF13taco},
imperatively~\cite{VisserPK04test,RuncimanNL08smallcheck,DanielDGM07automated,KurajKJ15programming}
or in a hybrid
fashion~\cite{RosnerBPKAFK14bounded,GligoricGJKKM10test}.

Unfortunately, these approaches are inefficient to automatically
leverage the invariant for the SPE problem. The key challenge of
adopting the existing enumeration techniques is to encode the
invariant. Specifically, the declarative enumeration techniques
specify the invariant and typically use generate-and-test
approaches. Our combinatorial SPE algorithm maintains the invariant of
the non-$\alpha$-equivalence. Let $P_1$ and $P_2$ be two programs of
the SPE set $\mathcal{P}$. The invariant is: $P_1 \ncong P_2$ for all
distinct $P_1, P_2 \in \mathcal{P}$. Therefore, to generate
$|\mathcal{P}|$ non-$\alpha$-equivalent programs, it needs to test
$\prod_{i=1}^{n} |\mathbb{v}_{i}|$ programs as a na\"{\i}ve SPE
solution discussed in Section~\ref{sec:ps}. In addition, the
imperative enumeration frameworks are capable of enumerating only
valid inputs \mywrt\ the invariant. However, our SPE algorithm solves
a combinatorial problem rather than generating combinatorial
structures (\myeg, red-black trees, graphs and algebraic
representations). Even though it might be feasible to encode the SPE
algorithm using the primitive enumerators in the imperative
enumeration frameworks, the realization is strictly less efficient
than directly applying our combinatorial SPE algorithm in the first
place.

Another relevant problem is enumerating lambda terms
exhaustively up to a given
size~\cite{GrygielL13counting,Lescanne13on,Tarau15on}. Most of the
work enumerates lambda terms using the standard ``nameless'' de
Bruijn representation~\cite{de_bruijn_indices}. These approaches
consider a rather different enumeration problem as the lambda terms
have distinct syntactic structures and semantics. Specifically, the
essential enumeration problem concerns with various unary-binary tree
structures~\cite{Lescanne13on,Tarau15on,GrygielL15counting}. However, in our set
partition setting, there is no dependence among set elements.

\paragraph{Algorithm correctness.}
Algorithm~\ref{algo:pe} invokes procedure \codeIn{PartitionScope} to
compute the scoped set partitions for each function $f$. We briefly
discuss the correctness of procedure \codeIn{PartitionScope}. Our algorithm
handles functions at
different granularities. In Algorithm~\ref{algo:pe}, the input
function $f$ is in the normal form. Recall that each hole $\square_i$
in the skeleton $\mathbb{P}$ corresponds to an element $i \in H$. In
the normal form, the elements can been filled with both global
($\mathbb{v}^g$) and local ($\mathbb{v}^l$) variables. We define the
\emph{configuration} of the normal form to be a map $c:H\to\{g,l\}$
for all variables $i\in H$. It is then sufficient and necessary to
show that: (1) procedure \codeIn{PartitionScope} computes unique
non-$\alpha$-equivalent partitions for each configuration; and (2)
procedure \codeIn{PartitionScope} finds all configurations in function
$f$.
\begin{itemize}
\item \emph{Part (1).} The configure $c$ maps $i\in H$ to either $l$
  and $g$, and it leads to two cases. In the first case, all elements are
  global. Therefore, the SPE problem becomes the standard set
  partition problem. Procedure \codeIn{PartitionScope} calls procedure
  \textsc{Partitions} to compute the set partitions of size $j$. In
  the second case, some elements $i$ representing local holes
  $\square^l$ are mapped to $l$. In this case, the partition problems
  of the global and local elements become independent. Procedure
  \codeIn{PartitionScope} computes respectively the set partitions for
  elements that representing both $\square^g$s and $\square^l$s, and
  obtains the global solution by computing their Cartesian product.

\item \emph{Part (2).} Procedure \codeIn{PartitionScope} calls
  procedure \textsc{Combinations} to find all configurations of function
  $f$'s normal form by exhaustively selecting the combinators of local
  holes.
\end{itemize}

\section{Evaluation}\label{sec:eval}
\begin{table*}[!t]
\footnotesize
\begin{center}
\begin{tabular}{ |c|rrr|rrr| } 
\hline
\multirow{2}{*}{Approach} & \multicolumn{3}{c|}{Original Test-Suite} & \multicolumn{3}{c|}{Enumerated Test-Suite}\\
 \cline{2-7}
 & Total Size & Avg. Size & \#Files & Total Size & Avg. Size & \#Files \\
\hline
Naive  & $5.24\times10^{163}$ & $2.49\times10^{159}$ & 20,978 & 1,310,943,547,383 & 69,538,698.7 & 18,852 \\
Our & $1.48\times10^{79}$ & $7.05\times 10^{74}$ & 20,978 & 2,050,671& 108.8 & 18,852 \\
 \hline
\end{tabular}
\end{center}
\caption{Evaluation results on size reduction. The ``total size''
  column shows the total numbers of enumerated programs, and the
  ``avg. size'' the average numbers of the enumerated programs for
  each test program. The size of the enumerated test-suite is related to a
  threshold discussed in Section~\ref{sec:expenum}.} \label{tab:pe}
\end{table*}

To evaluate the effectiveness of skeletal program enumeration, we
conduct two sets of experiments. In the first experiment, we enumerate
skeletons derived from GCC-4.8.5's test-suite, and test two stable
compiler releases, GCC-4.8.5 and Clang-3.6.1. We aim to demonstrate
the benefits of combinatorial SPE. In the second experiment, we use a set of small programs to test the
trunk versions of GCC and Clang, as well as CompCert and two Scala compilers, to demonstrate the bug-hunting
capabilities of SPE.

\subsection{Experimental Setup}\label{sec:expsetup}

Our implementation contains two
components, \myie, skeleton generation and program enumeration. The
skeleton generation component recursively traverses the ASTs to obtain the scope and type
information for each variable, and build a skeleton $\mathbb{P}$ for
each test program $P$. The program enumeration component realizes the enumeration algorithm
described in Algorithm~\ref{algo:pe}. We compute the intra-procedural enumeration as mentioned in
Section~\ref{sec:pediss}.

Given a set of programs $\mathcal{P}$, we directly feed those
programs to the compilers under testing. For GCC and Clang, we use two optimization
levels (\myie, -O0 and -O3) and two machine modes (\myie, 32- and
64-bits) for finding crashes. For wrong code bugs, we investigate the program $P$ with CompCert's reference
interpreter~\cite{Leroy06formal} and additional manual efforts to ensure that it
is free of undefined behaviors. All experiments were conducted on a
server and a desktop running Ubuntu-14.04. The server has Intel Xeon
X7542 CPUs and 128GB RAM, while the desktop has an Intel i7-4770 CPU and
16GB RAM.

\subsection{Experiments on Stable Releases}\label{sec:expstable}

In our first experiment, we evaluate the SPE technique on stable
releases of two popular C compilers, specifically GCC-4.8.5 and
Clang-3.6.1. We choose GCC-4.8.5 since it is the default C compiler in
the long term support version of Ubuntu (14.04), and Clang-3.6.1 was
released about the same time as the chosen GCC.

We implemented both our combinatorial program enumeration described in
Algorithm~\ref{algo:pe} and a na\"{\i}ve enumeration algorithm mentioned in
Section~\ref{sec:ps}. We apply the two implementations on the default
test-suite which has been shipped with GCC-4.8.5. Most of the test
programs belong to the c-torture suite, which contains the code
snippets that have historically broken previous
releases.\footnote{\url{https://gcc.gnu.org/onlinedocs/gccint/C-Tests.html}.}
According to the GCC's release criteria, any released version must
pass the test-suite distributed in the source code. We are
particularly interested in understanding the following research
questions: 
\begin{itemize}
\item What is the size reduction achieved by our SPE
  approach?

\item Given the fact that the test-suite contains many programs once
  broke previous GCCs, what are the characteristics of these programs?

\item Can SPE find bugs in the stable GCC and Clang releases using their own regression test-suite?
\end{itemize}

\subsubsection{Enumeration Size Reduction}\label{sec:expenum}

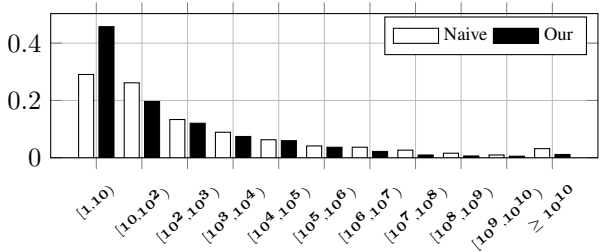
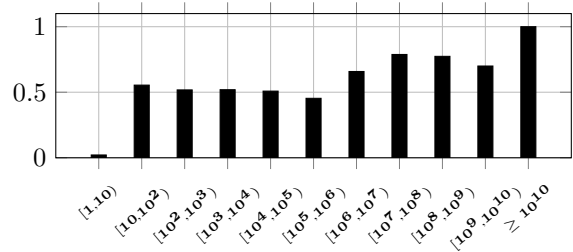
\begin{figure*}[!t]
\centering

\subfigure[Distribution of the numbers of variants. For example, the
  first pair of vertical bars shows that 29\% of the test programs
  have fewer than 10 variants enumerated using the na\"{\i}ve approach. The
  percentage increases to 46\% using our approach.]{
  \begin{tikzpicture}
\begin{axis}[
  xtick=data,
  ymin=0,
  bar width=.2cm,
  ybar,
  height=3.5cm,
  width=0.5\textwidth,
  grid=major,
  x tick label style={rotate=40, font=\tiny},  
  legend style={anchor=north east,legend columns=0,font=\scriptsize},
  area legend,
  xticklabels={$\mathbf{[1\mbox{,} 10)}$,
      $\mathbf{[10\mbox{,} 10^2)}$,
      $\mathbf{[10^2\mbox{,} 10^3)}$,      
      $\mathbf{[10^3\mbox{,} 10^4)}$,
      $\mathbf{[10^4\mbox{,} 10^5)}$,      
      $\mathbf{[10^5\mbox{,} 10^6)}$,      
      $\mathbf{[10^6\mbox{,} 10^7)}$,  
      $\mathbf{[10^7\mbox{,} 10^8)}$,      
      $\mathbf{[10^8\mbox{,} 10^9)}$, 
      $\mathbf{[10^9\mbox{,} 10^{10})}$,      
      $\mathbf{\ge 10^{10}}$, 
  }
]

  \addplot [black, fill=white] table [x expr=\coordindex, y index=1] {data/processed-num-of-variants-table.txt};
  \addplot [black, fill=black] table [x expr=\coordindex, y index=2] {data/processed-num-of-variants-table.txt};
\legend{Naive, Our} 
\end{axis} 
\end{tikzpicture}  }\hspace{0.2cm}
\subfigure[Distribution of the ratios of reduced variants.  For
  example, the second bar shows that our approach has eliminated 55\%
  of the programs with $n \in [10,10^2)$ variants compared with the
    na\"{\i}ve approach on average. ]{ \begin{tikzpicture}

\begin{axis}[
  xtick=data,
  ymin=0,
  ybar,
  height=3.5cm,
  width=\columnwidth,
  bar width=.2cm,  
  grid=major,
  x tick label style={rotate=40, font=\tiny},  
  legend style={anchor=north east,legend columns=0},
  xticklabels={$\mathbf{[1\mbox{,} 10)}$,
      $\mathbf{[10\mbox{,} 10^2)}$,
      $\mathbf{[10^2\mbox{,} 10^3)}$,      
      $\mathbf{[10^3\mbox{,} 10^4)}$,
      $\mathbf{[10^4\mbox{,} 10^5)}$,      
      $\mathbf{[10^5\mbox{,} 10^6)}$,      
      $\mathbf{[10^6\mbox{,} 10^7)}$,  
      $\mathbf{[10^7\mbox{,} 10^8)}$,      
      $\mathbf{[10^8\mbox{,} 10^9)}$, 
      $\mathbf{[10^9\mbox{,} 10^{10})}$,      
      $\mathbf{\ge 10^{10}}$, 
  }
]

  \addplot [fill=black] table [x expr=\coordindex, y index=1] {data/processed-size-improvement-by-reduction.txt};
\end{axis} 
\end{tikzpicture}  }
\caption{Overview of the size reduction. In both figures, the $x$-axis
  lists the size ranges of enumeration sets $\mathcal{P}$.  In
  particular, $\mathcal{P}$ is described using the number of variants
  enumerated for each test program $P$. The $y$-axis represents the
  percentage.}\label{fig:pedis}
\end{figure*}

The GCC-4.8.5 test-suite contains about 21K C
files. Table~\ref{tab:pe} decries the size reduction results of
applying our combinatorial SPE algorithm. For the original test-suite,
our combinatorial SPE approach reduces the entire size by 94 orders of
magnitude. However, it is clear from the table that SPE solution set
is still too large to be applied for compiler testing in practice. As
a result, we set a 10K threshold such that we
ignore the test programs which have more than 10K variants using our
combinatorial SPE algorithm.
The 10K threshold is chosen \mywrt\ the characteristics of the
test-suite (Table~\ref{tab:g485}), \myie, ${\mathit{|Vars|}}^{\mathit{|Holes|}} = 3.46^{7.34} \approx 10K$).
We then compare the solution spaces based
on the remaining programs. From the last three columns in
Table~\ref{tab:pe}, we can see that the number of test files is
decreased to 19K. Using the 10K threshold, we can still retain
90\% of the original test programs. On those files, our
SPE algorithm achieves six orders of size
reduction over the na\"{\i}ve approach. Specifically, for each test
program, the solution of our SPE approach contains merely 109
files on average. In practical settings, suppose that we could process
each program in one second, it takes less than one month to
handle all enumerated programs. However, for the na\"{\i}ve approach, it
takes about 40K years to process the same test programs. Finally,
Figure~\ref{fig:pedis} describes size reduction in terms of different
program enumeration sets $\mathcal{P}$.

\subsubsection{Test-Suite Characteristics}

Table~\ref{tab:g485} gives an overview of the test programs in
GCC-4.8.5 test-suite. It also describes the programs used in our
evaluation based on the aforementioned 10K threshold. It is
interesting to observe that most of the programs are quite small even
though most of them have triggered bugs in previous versions of
GCC. Indeed, this observation has motivated our current program
enumeration work. The programs used in our evaluation are smaller due
to the 10K threshold setting \mywrt\ our combinatorial enumeration
algorithm. Recall that these programs represent 90\% of the programs
in the original test-suite. It clearly demonstrates that it is
feasible to apply combinatorial SPE on
practical test-suites.

\subsubsection{Benefits of Skeletal Program Enumeration}\label{sec:bene}

\begin{table}[!t]
\footnotesize
\begin{center}
\begin{tabular}{ |c|rrrrr| } 
\hline
Test-Suite&\#Holes&\#Scopes&\#Funcs&\#Types&\#Vars\\
\hline

Original & 7.34& 2.77 &1.85 &1.38 &3.46 \\
Enumerated & 3.84 & 1.85&1.50 &1.29 & 1.60\\
 \hline
\end{tabular}

\end{center}
\caption{Characteristics of the GCC-4.8.5 test-suite. The  first four columns display the average counts of holes, scopes,
  functions and variable types in each file, respectively. The last
  column displays the variable counts for each hole.} \label{tab:g485}
\end{table}



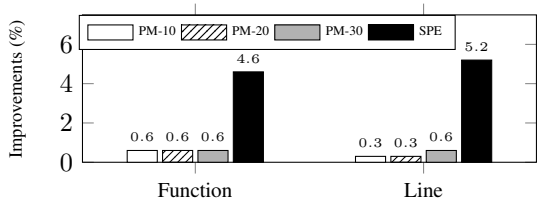
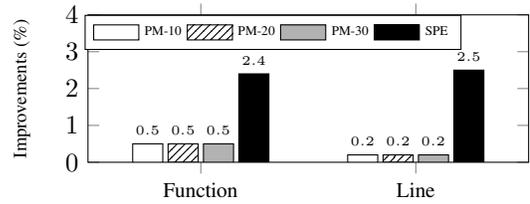
\begin{figure*}[!t]
\centering

\subfigure[GCC-4.8.5 coverage improvements.]{
\begin{tikzpicture}
    \begin{axis}[
            ybar,   enlarge x limits={abs=1.5cm},
    symbolic x coords={Function, Line},
    xtick=data,
bar width=.4cm,
nodes near coords,
nodes near coords align={vertical},
every node near coord/.append style={font=\tiny},
            ymin=0,width=0.43\textwidth,ymax=7.5,
x tick label style={font=\footnotesize},
ylabel= {Improvements (\%)},
y label style={at={(axis description cs:0.1,.5)},anchor=south, font=\scriptsize},
            height=.2\textwidth,
legend style={at={(0.407,1)}, anchor=north,legend columns=0, font=\tiny},
  area legend
        ]
\addplot [black,fill=black!0] coordinates {(Function,0.6) (Line,0.3)};
\addplot [black,fill=black!0,pattern=north east lines]coordinates {(Function,0.6) (Line,0.3)};
\addplot [black,fill=black!30]coordinates {(Function,0.6) (Line,0.6)};
\addplot [black,fill=black!100] coordinates {(Function,4.6) (Line,5.2)};
\legend{PM-10, PM-20,PM-30,SPE}

    \end{axis}
\end{tikzpicture}

} \hspace{2cm}
\subfigure[Clang-3.6 coverage improvements.]{
\begin{tikzpicture}
    \begin{axis}[
            ybar,   enlarge x limits={abs=1.5cm},
    symbolic x coords={Function, Line},
    xtick=data,
bar width=.4cm,
nodes near coords,
nodes near coords align={vertical},
every node near coord/.append style={font=\tiny},
            ymin=0,width=0.42\textwidth,ymax=4,
x tick label style={font=\footnotesize},
ylabel= {Improvements (\%)},
y label style={at={(axis description cs:0.1,.5)},anchor=south, font=\scriptsize},
            height=.2\textwidth,
legend style={at={(0.42,1)}, anchor=north,legend columns=0, font=\tiny},
  area legend
        ]
\addplot [black,fill=black!0] coordinates {(Function,0.5) (Line,0.2)};
\addplot [black,fill=black!0,pattern=north east lines]coordinates {(Function,0.5) (Line,0.2)};
\addplot [black,fill=black!30]coordinates {(Function,0.5) (Line,0.2)};
\addplot [black,fill=black!100] coordinates {(Function,2.4) (Line,2.5)};
\legend{PM-10, PM-20,PM-30,SPE}

    \end{axis}
\end{tikzpicture}

}
\caption{Coverage improvements over the baseline tests. PM-X represent
  improvements achieved by program mutation (the \textsf{Orion} tool)
  which deletes X statements, and SPE represents improvements achieved
  using our SPE algorithm.}\label{fig:comppm}
\end{figure*}


\paragraph{Hunting bugs.}
We apply our enumeration algorithm to test GCC-4.8.5 and
Clang-3.6 by enumerating the skeletons from the GCC-4.8.5
test-suite. Our SPE technique have found 1 and 10 crash bugs in GCC and Clang, respectively.
It is perhaps interesting to note that we are able to find GCC
bugs by enumerating its own test-suite even if the release criteria
force it to pass the original test-suite.  In this evaluation, we only
focus on crash bugs since wrong code bugs usually require compiler
developers' confirmation (mostly due to possible undefined behaviors
in test programs). For crash bugs, compiler messages clearly indicates
their occurrence.  Table~\ref{tab:bugsig} gives the signatures of some
crash bugs found in this evaluation. We can see that most of the bugs
are in the backend and optimization passes.

\begin{table}[!t]
\scriptsize
\begin{center}
\begin{tabular}{ |l| } 
\hline
internal compiler error: in assign\_by\_spills, at lra-assigns.c:1281\\
\hline
error in backend: Do not know how to split the result of this
operator!\\
error in backend: Invalid register name global variable.\\
error in backend: Access past stack top!\\
Assertion `MRI->getVRegDef(reg) \&\& ``Register use before def!''' failed.\\
Assertion `Num < NumOperands \&\& ``Invalid child \# of SDNode!''' failed.\\
 \hline
\end{tabular}
\end{center}

\caption{Crash signatures of bugs found in GCC-4.8.5 and Clang-3.6.1 using the GCC-4.8.5 test-suite.}\label{tab:bugsig}
\end{table}

\paragraph{Improving coverage.}
As described in Section~\ref{sec:intro}, one of our insights is that
SPE can help trigger more internal compiler passes.  In order to
validate the claim, we compare our SPE technique against the seminal
work \textsf{Orion} of program mutation~\cite{LeAS14compiler}. We
choose \textsf{Orion} since it only considers statement
deletion. Therefore, the overall search spaces for both approaches are
bounded. We randomly select 100 test programs from the test-suite to
run both approaches. Figure~\ref{fig:comppm} gives the empirical
results. The selected test programs achieve 41\% function coverage
and 32\% line coverage for GCC, respectively. For Clang, they
achieve 20\% function coverage and 17\% line coverage,
respectively. Our SPE approach brings approximately 5\% coverage
improvement for GCC and 2.4\% coverage improvement for Clang,
respectively. On the other hand, \textsf{Orion} provides less than 1\%
coverage improvement. This comparison also demonstrates the advantage
of applying our SPE technique on small programs.

It is also worth noting that \textsf{Orion} has found 1 and 3 bugs in
Clang-3.6 and GCC-4.8.5, respectively, using the same 
test-suite. The three GCC bugs are unique as they are different from
what we have found. This evaluation has also provided practical
evidence that program enumeration and program mutation offer 
complementary benefits.

\subsection{Experiments on Development Versions}\label{sec:exptrunk}

We apply our combinatorial program enumeration for finding bugs in the
trunk versions of GCC and Clang. We select a set of small C programs
from the unit test-suite of many open-source projects, such as
CompCert~\cite{Leroy06formal}, Frama-C,
the Rose compiler and
KCC~\cite{EllisonR12an}. In particular, most of our test C programs
are from the test-suite in the trunk version of GCC. The test programs
share similar characteristics with those described in
Section~\ref{sec:expstable}. We began our testing process in early
January. In less than six months, our technique has discovered 217  GCC and
Clang bugs. To date, more than half of them have
been fixed. 

To demonstrate SPE's generality, we have also applied it to test the CompCert
verified C compiler, and two optimizing Scala compilers, \myie, the
production Scala compiler and the Dotty research compiler. In about
three weeks, we have reported 29 CompCert crashing bugs and 42 bugs in the
two Scala compilers. The developers have appreciated and
promptly addressed our reports --- 25 CompCert bugs have already been
fixed (all have been confirmed), and 27 Dotty bugs have been
confirmed. We started testing the two Scala compilers recently in late October. Among the Dotty bugs, 9 have been fixed so far. 
Until now, there are only five
high-priority bugs in total in the Dotty code repository, and
our SPE technique has discovered four of them.
The rest of this section focuses discussing the GCC and Clang/LLVM bugs.

\subsubsection{Overall Results}

\begin{table*}[!t]
\footnotesize
\begin{center}
\begin{tabular}{ |c|rrrrr|rrr| } 
\hline
\multirow{2}{*}{Compiler} & \multicolumn{5}{c|}{Summary} & \multicolumn{3}{c|}{Classification}\\
 \cline{2-9}
& Reported & Fixed & Duplicate & Invalid & Reopened & Crash & Wrong code & Performance\\
\hline
GCC& 136 & 93 & 10 & 2 & 1 & 127 & 6 & 3\\
Clang& 81 & 26 & 3 & 1 & 1 & 79 & 2 & 0\\

 \hline
\end{tabular}
\end{center}
\caption{Overview of bugs reported for trunk versions of GCC and Clang in six months.} \label{tab:over}
\end{table*}
\pgfplotstableread[row sep=\\,col sep=&]{
    interval & Rep & Fix \\
    P1     & 13  & 12  \\
    P2     & 39 & 27  \\
    P3    & 74 & 48 \\
    P4-5   & 10 & 6 \\
    }\gccpio

\pgfplotstableread[row sep=\\,col sep=&]{
    interval & Rep & Fix \\
    -O0     & 77  & 47  \\
    -O1     & 25 & 17  \\
    -O2    & 40 & 30 \\
    -O3   & 51 & 39 \\
    }\gccopt

\pgfplotstableread[row sep=\\,col sep=&]{
    interval & Rep & Fix \\
    Earlier     & 58  & 36  \\
    5.X     & 90 & 62  \\
    6.X    & 116 & 79 \\
    Trunk   & 136 & 94 \\
    }\gccver

\pgfplotstableread[row sep=\\,col sep=&]{
    interval & Rep & Fix \\
    C     & 13  & 11  \\
    C++     & 63 & 37  \\
    Inline-asm    & 2 & 0 \\
    IPA   & 2 & 1 \\
    Middle-end   & 10 & 7 \\
    RTL-optimization & 6 & 4\\
    Target & 6 & 5\\
    Tree-optimization & 34 & 28\\
}\gcccomp

\begin{figure*}[!t]
\centering
\subfigure[Bug priorities.]{
\begin{tikzpicture}
    \begin{axis}[
            ybar,
bar width=.2cm,
nodes near coords,
            symbolic x coords={P1,P2,P3,P4-5},
            xtick=data,ymax=95,
nodes near coords align={vertical},
every node near coord/.append style={font=\tiny},
            ymin=0,width=0.3\textwidth,x tick label style={font=\footnotesize},
            height=.18\textwidth,
        ]
        \addplot[black,fill=black!30] table[x=interval,y=Fix]{\gccpio};
        \addplot[black,fill=black!80] table[x=interval,y=Rep]{\gccpio};
    \end{axis}
\end{tikzpicture}
\label{fig:pios}
}\hspace{1cm}
\subfigure[Affected optimization levels.]{
\begin{tikzpicture}
    \begin{axis}[
            ybar,
bar width=.2cm,
nodes near coords,
            symbolic x coords={-O0,-O1,-O2,-O3},
            xtick=data,ymax=95,
nodes near coords align={vertical},
every node near coord/.append style={font=\tiny},
            ymin=0,width=0.3\textwidth,
x tick label style={font=\footnotesize},
            height=.18\textwidth,
        ]
        \addplot[black,fill=black!30] table[x=interval,y=Fix]{\gccopt};
        \addplot[black,fill=black!80] table[x=interval,y=Rep]{\gccopt};
    \end{axis}
\end{tikzpicture}
\label{fig:opts}
}\hspace{1cm}
\subfigure[Affected GCC versions.]{
\begin{tikzpicture}
    \begin{axis}[
            ybar,
bar width=.2cm,
nodes near coords, ymax=165,
            symbolic x coords={Earlier,5.X,6.X,Trunk},
            xtick=data,x tick label style={font=\footnotesize},
nodes near coords align={vertical},
every node near coord/.append style={font=\tiny},
            ymin=0,width=0.3\textwidth,
            height=.18\textwidth,
        ]
        \addplot[black,fill=black!30] table[x=interval,y=Fix]{\gccver};
        \addplot[black,fill=black!80] table[x=interval,y=Rep]{\gccver};
    \end{axis}
\end{tikzpicture}
\label{fig:vers}
}
\subfigure[Affected GCC components.]{
  \begin{tikzpicture}

    \begin{axis}[
            ybar,
bar width=.3cm,
nodes near coords,
            symbolic x coords={C,C++,Inline-asm,IPA,Middle-end,RTL-optimization,Target,Tree-optimization},
            xtick=data,
nodes near coords align={vertical},
every node near coord/.append style={font=\tiny},
x tick label style={rotate=10,font=\footnotesize},
            ymin=0,width=0.9\textwidth, ymax=75,
            height=.18\textwidth,
        ]
        \addplot[black,fill=black!30] table[x=interval,y=Fix]{\gcccomp};
        \addplot[black,fill=black!80] table[x=interval,y=Rep]{\gcccomp};
    \end{axis}
\end{tikzpicture}
\label{fig:comps}
}
\caption{Characteristics of GCC trunk bugs. The darker bars denote the
  numbers of reported bugs and the lighter bars the numbers of fixed
  bugs.} \label{fig:bugchara}
\end{figure*}
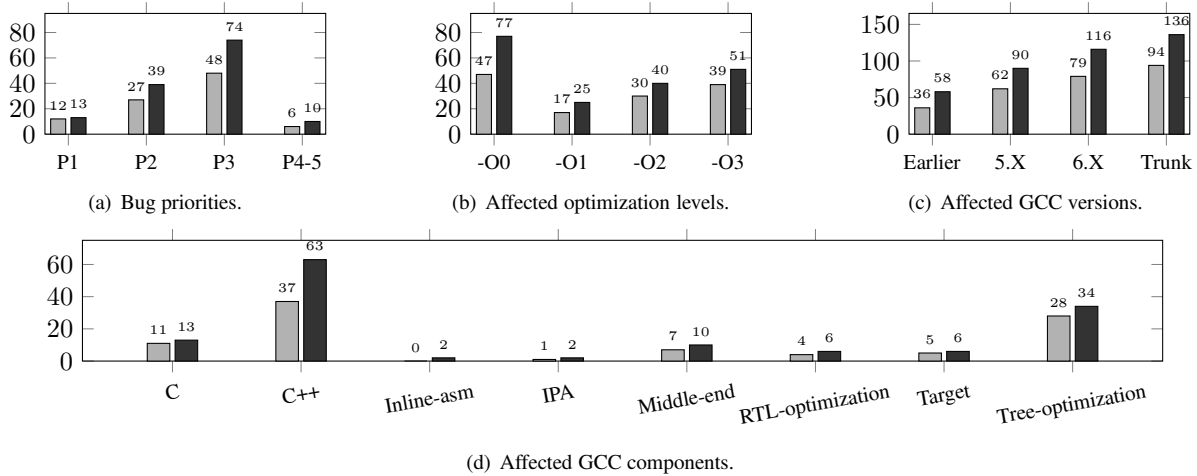

Table~\ref{tab:over} gives an overview of the bugs that we have found
during the testing course. We have reported 217 bugs in
total. Developers have confirmed almost all of our reported
bugs. Moreover, more than half of them have already been fixed within
the six-month period. Some of our reported bugs are quite complex. For
example, two bugs have been reopened by developers for further
inspection. Although we ensure that our reported bugs have different
symptoms, it is sometimes inevitable that we have occasionally reported
duplicates as it is quite difficult for us to track the root cause for
each bug. However, less than 5\% of the bugs are duplicates. Two of
our reported GCC bugs have been marked as invalid.  In particular, one
of them is about multiple inheritance and casting in C++, and the
other is a C program that contains undefined behavior concerned with
strict aliasing. We further discuss the undefined behavior issue in
Section~\ref{sec:ubdis}.

Table~\ref{tab:over} also gives the classification of the bugs. Most
of the bugs cause compiler crashes. As mentioned in
Section~\ref{sec:expsetup}, we leverage the SPE technique to find both
frontend and optimization bugs. Among all GCC crash bugs, 56\% of
them trigger frontend crashes, where most of them are related to the
C++ frontend. On the other hand, 44\% lead to crashes in the
optimization passes. Moreover, we have discovered 8 bugs related to
miscompilation. As mentioned in Section~\ref{sec:moti}, one of
them has been around for more than ten years. Finally, three of the
bugs are related to compilation performance. We describe one such
bug in \ifapps Appendix~\ref{app:additional}. \else the supplementary material. \fi

\newsavebox\myboxGccSampleOne
\begin{lrbox}{\myboxGccSampleOne}
  \begin{minipage}{.24\textwidth}%
    \lstset{language=C, 
      mathescape, 
      basicstyle=\scriptsize\ttfamily,  
      keywordstyle=\bfseries,   
      numbers=left, 
      xleftmargin=0.8cm,  
      numberstyle=\tiny,  
      numbersep=5pt,     
      showstringspaces=false,  
      upquote=true,  
      tabsize=2}
    \renewcommand{\ttdefault}{pcr}
    \begin{lstlisting}
class A {
  virtual void foo()
  { }
};

class B : public A, A
{ };

B b1, &b2 = b1;
A a = b2;
    \end{lstlisting}
  \end{minipage}
\end{lrbox}

\newsavebox\myboxGccSampleTwo
\begin{lrbox}{\myboxGccSampleTwo}
  \begin{minipage}{.3\textwidth}%
    \lstset{language=C, mathescape, basicstyle=\scriptsize\ttfamily,  
      keywordstyle=\bfseries,    
      xleftmargin=0.15cm,  numberstyle=\tiny,  
      numbersep=5pt,     showstringspaces=false,  
      upquote=true,  tabsize=2, numbers=left}
    \renewcommand{\ttdefault}{pcr}
    \begin{lstlisting}
void foo()
{
  unsigned long l;
  void *p = 0; 
  
  __builtin_unwind_init ();
  
  l = 0;
  
  __builtin_eh_return (l, p);
}
    \end{lstlisting}
\vspace*{.6cm}
  \end{minipage}
\end{lrbox}

\newsavebox\myboxGccSampleThree
\begin{lrbox}{\myboxGccSampleThree}
  \begin{minipage}{.20\textwidth}%
    \lstset{language=C, 
      mathescape, basicstyle=\scriptsize\ttfamily,  
      keywordstyle=\bfseries,    xleftmargin=0.5cm,  
      numberstyle=\tiny,  numbersep=5pt,     
      showstringspaces=false,  upquote=true,  
      tabsize=2, numbers=left}
    \renewcommand{\ttdefault}{pcr}
    \begin{lstlisting}
char a; short b;
void fn1() {
  if (b)
    ;
  else {
    int c[1] = {0};
    l1: ;
  }
  if (a) goto l1;
}
    \end{lstlisting}
  \end{minipage}
\end{lrbox}

\newsavebox\myboxGccSampleFour
\begin{lrbox}{\myboxGccSampleFour}
  \begin{minipage}{.3\textwidth}%
    \lstset{language=C, mathescape, 
      basicstyle=\scriptsize\ttfamily,  keywordstyle=\bfseries,    
      xleftmargin=0.15cm,  numberstyle=\tiny,  
      numbersep=5pt,     showstringspaces=false,  
      upquote=true,  tabsize=2, numbers=left}
    \renewcommand{\ttdefault}{pcr}
    \begin{lstlisting}
double u[1782225];
int a, b, d, e;
static void foo(int *p1) {
  double c = 0.0;
  for (; a < 1335; a++) {
    b = 0;
    for (; b < 1335; b++)
      c = c + u[a + 1335 * a];(*@\label{line:gcc:70318:replacement}@*)
    u[1336 * a] *= 2;
  }
  *p1 = c;
}
int main() {...}
    \end{lstlisting}
  \end{minipage}
\end{lrbox}

\newsavebox\myboxGccSampleFive
\begin{lrbox}{\myboxGccSampleFive}
  \begin{minipage}{.24\textwidth}%
    \lstset{language=C, mathescape, 
      basicstyle=\scriptsize\ttfamily,  keywordstyle=\bfseries,    
      xleftmargin=0.6cm,  numberstyle=\tiny,  
      numbersep=5pt,     showstringspaces=false,  
      upquote=true,  tabsize=2, numbers=left}
    \renewcommand{\ttdefault}{pcr}
    \begin{lstlisting}
int a;
double b, *c;

void fn1(int p1) {
  for (;; p1--) {
    a = p1;
    for (; p1 >= a; a--)
      b = c[p1];
  }
}
    \end{lstlisting}
  \end{minipage}
\end{lrbox}

\newsavebox\myboxGccSampleSix
\begin{lrbox}{\myboxGccSampleSix}
  \begin{minipage}{.25\textwidth}%
    \lstset{language=C, mathescape, 
      basicstyle=\scriptsize\ttfamily,  keywordstyle=\bfseries,    
      xleftmargin=1.1cm,  numberstyle=\tiny,  
      numbersep=5pt,     showstringspaces=false,  
      upquote=true,  tabsize=2, numbers=left}
    \renewcommand{\ttdefault}{pcr}
    \begin{lstlisting}
int main() {
  int *p = 0;
trick:
  if (p)
    return *p;
  int x = 0;
  p = &x;
  goto trick;
  return 0;
}
    \end{lstlisting}
  \end{minipage}
\end{lrbox}

\newsavebox\myboxGccSampleSeven
\begin{lrbox}{\myboxGccSampleSeven}
	\begin{minipage}{.3\textwidth}%
		\lstset{language=C++, mathescape, 
			basicstyle=\scriptsize\ttfamily,  keywordstyle=\bfseries,    
			xleftmargin=0.15cm,  numberstyle=\tiny,  
			numbersep=5pt,     showstringspaces=false,  
			upquote=true,  tabsize=2, numbers=left}
		\renewcommand{\ttdefault}{pcr}
		\begin{lstlisting}
struct C {
  C() {}
  int i;
};

void *operator new(size_t, void *p2)  (*@\label{line:gcc:71405:replacement}@*)
{ return p2; }

int main() {
  int a;
  new (&a) C;
  return 0; 
}
		\end{lstlisting}
	\end{minipage}
\end{lrbox}

\newsavebox\myboxClangSampleEight
\begin{lrbox}{\myboxClangSampleEight}
	\begin{minipage}{.3\textwidth}%
		\lstset{language=C++, mathescape, 
			basicstyle=\scriptsize\ttfamily,  keywordstyle=\bfseries,    
			xleftmargin=0.15cm,  numberstyle=\tiny,  
			numbersep=5pt,     showstringspaces=false,  
			upquote=true,  tabsize=2, numbers=left}
		\renewcommand{\ttdefault}{pcr}
		\begin{lstlisting}
enum Color { 
  R, G, B 
};	
	
template < typename T >
void test(T, __underlying_type (T)) (*@\label{line:clang:28045:test}@*) 
{}

int main() { 
  Color c = R;
  test (c, c);
  return 0;
}
		\end{lstlisting}
	\end{minipage}
\end{lrbox}

\newsavebox\myboxCcompSampleOne
\begin{lrbox}{\myboxCcompSampleOne}
	\begin{minipage}{.25\textwidth}%
		\lstset{language=C, mathescape, 
			basicstyle=\scriptsize\ttfamily,  keywordstyle=\bfseries,    
			xleftmargin=0.15cm,  numberstyle=\tiny,  
			numbersep=5pt,     showstringspaces=false,  
			upquote=true,  tabsize=2, numbers=left}
		\renewcommand{\ttdefault}{pcr}
		\begin{lstlisting}
union U u = { 0 };
		\end{lstlisting}
	\end{minipage}
\end{lrbox}

\newsavebox\myboxCcompSampleTwo
\begin{lrbox}{\myboxCcompSampleTwo}
	\begin{minipage}{.28\textwidth}%
		\lstset{language=C, mathescape, 
			basicstyle=\scriptsize\ttfamily,  keywordstyle=\bfseries,    
			xleftmargin=0.15cm,  numberstyle=\tiny,  
			numbersep=5pt,     showstringspaces=false,  
			upquote=true,  tabsize=2, numbers=left}
		\renewcommand{\ttdefault}{pcr}
		\begin{lstlisting}
void foo (struct A a) 
{
  a++;
}
		\end{lstlisting}
	\end{minipage}
\end{lrbox}

\newsavebox\myboxScalaSampleOne
\begin{lrbox}{\myboxScalaSampleOne}
	\begin{minipage}{.28\textwidth}%
		\lstset{language=scala, mathescape, 
			basicstyle=\scriptsize\ttfamily,  keywordstyle=\bfseries,    
			xleftmargin=0.15cm,  numberstyle=\tiny,  
			numbersep=5pt,     showstringspaces=false,  
			upquote=true,  tabsize=2, numbers=left}
		\renewcommand{\ttdefault}{pcr}
		\begin{lstlisting}
class Bar {
  def f (x : { def g }) {}
  f (new Foo { def g })
}
		\end{lstlisting}
	\end{minipage}
\end{lrbox}

\newsavebox\myboxDottySampleTwo
\begin{lrbox}{\myboxDottySampleTwo}
	\begin{minipage}{.25\textwidth}%
		\lstset{language=scala, mathescape, 
			basicstyle=\scriptsize\ttfamily,  keywordstyle=\bfseries,    
			xleftmargin=0.15cm,  numberstyle=\tiny,  
			numbersep=5pt,     showstringspaces=false,  
			upquote=true,  tabsize=2, numbers=left}
		\renewcommand{\ttdefault}{pcr}
		\begin{lstlisting}
object Main extends App {
  case class Foo(field: Option[String])
  val x: PartialFunction[Foo, Int] = { 
    c => c.field match {
      case Some(s) => 42
    }
  }
}
		\end{lstlisting}
	\end{minipage}
\end{lrbox}

\begin{figure*}[!t]
\centering
\subfigure[G++ crash \anonybugid{70202} (fixed)]{
  \usebox\myboxGccSampleOne
\label{subfig:sample:gcc:70202}
} \hspace{.0cm}
\subfigure[\gcc crash \anonybugid{69740} (reopened)]{
  \usebox\myboxGccSampleThree
\label{subfig:sample:gcc:69740}
}\hspace{.0cm}
\subfigure[\clang crash \anonybugid{26973} (fixed)]{
  \usebox\myboxGccSampleFive
  \label{subfig:sample:clang:26973}
}\hspace{.0cm}
\subfigure[\clang wrong code \anonybugid{26994} (confirmed)]{
  \usebox\myboxGccSampleSix
  \label{subfig:sample:26994}
}
\vspace*{-3pt}
\caption{Sample test programs that trigger bugs of \gcc and \clang.} \label{fig:bugtrunkchara}
\end{figure*}

\subsubsection{Bug Characteristics}

We discuss the characteristics of our reported GCC
bugs. It is worth mentioning that we have made more effort testing GCC
since GCC developers are relatively more responsive. In particular,
GCC developers not only have fixed 68\% of our bugs but also provide
more feedback. Figure~\ref{fig:bugchara} characterizes the 136
reported GCC trunk bugs. Specifically, Figure~\ref{fig:pios} shows the importance of the reported  bugs. P3 is the default priority in GCC's bugzilla system. About two
  thirds of the bugs fall into this category. About 10\% of them are
  release-blocking (P1). Developers have to fix all P1 bugs
  in order to release a future version. Figure~\ref{fig:opts} shows that our reported bugs cover all
  optimization levels. Specifically, our approach has found more -O3
  bugs than the -O2 and -O1 bugs. This demonstrates that the SPE
  technique is able to cover deep compiler optimization passes. Figure~\ref{fig:vers} shows the affected GCC versions. We can
  see that 85\% of the bugs affect the latest 6 release. Moreover,
  66\% of the bugs affect at least three stable GCC 5
  releases. Perhaps the most interesting to note is that 43\%
  of the bugs affect earlier GCC versions from at least one year
  ago. It demonstrates that our techniques can find long latent
  bugs. Figure~\ref{fig:comps} shows the diversity of our reported
  bugs. Over half of our bugs are C++ frontend bugs. The second
  category of most frequent bugs concern the tree-optimization
  component. The results suggest that our SPE technique is useful for
  testing various compiler components.

Our technique has discovered a large number of diverse bugs in a
relatively short period of time.  One unique, noteworthy aspect of our
work is the large number of reported bugs in the compilers' C++
support, making it the first successful exhaustive technique to provide
this capability.  C++ is an active, enormously complex language and
has a growing set of features --- it is very challenging to develop
practical C++ program generators.  Note that we have many more bugs to
triage, reduce and report, but have been reporting bugs in a steady
fashion so as not to overwhelm the developers.  The results highlight
the novelty and benefits of our approach.

\subsubsection{Case Studies on Sample Bugs}\label{sec:cnbugs}
We select and discuss four reported \gcc and \clang
bugs. Figure~\ref{fig:bugtrunkchara} describes the corresponding test
programs with bug classifications and status. Eight additional bug samples 
may be found in  \ifapps Appendix~\ref{app:additional}. \else the anonymized supplementary material. \fi


\paragraph{Figure~\ref{subfig:sample:gcc:70202}.} This test program exposes a long latent bug of \gcc that affects all
versions since \gcc-4.4, which was released four years ago.  The bug is
in the C++ frontend of \gcc, and manifests when \gcc computes the path
of the base classes for the class \codeIn{B}.  The GCC developers have
confirmed this bug and are investigating its root cause.

\paragraph{Figure~\ref{subfig:sample:gcc:69740}.}
This is a crash bug of the \gcc trunk (6.0 revision 233242).  It
manifests when \gcc compiles the test program at -O2 and above.
The \codeIn{goto} statement in the program introduces an irreducible
loop, and \gcc incorrectly handles the backend and consequently
triggers the assertion \codeIn{verify\_loop\_structure} to fail.  This
reported bug had been fixed once, and later reopened by the GCC
developers.  Note that this program is enumerated from the test
program in \gcc bug report PR\anonybugid{68841}.

\paragraph{Figure~\ref{subfig:sample:clang:26973}.}
This test program crashes the trunk (3.9 revision 263641) of \clang at
-O1 and above.  This bug is a regression, and had been latent for
eleven months until we discovered it.  The culprit revision incorrectly
passes a wrong parameter to infer the loop invariant, and consequently
corrupts the emitted LLVM bitcode and causes an assertion violation in
the compiler backend.

\paragraph{Figure~\ref{subfig:sample:26994}.}
The program is miscompiled by the \clang trunk (3.9.0 revision
263789).  The expected exit code is 0. However the miscompiled
executable returns 1 instead.  The root cause is that the \clang
frontend deems that the lifetime of the variable \codeIn{x} ends after
the control flow jumps to the label \codeIn{trick}, which is
incorrect.  Consequently the write to variable \codeIn{x}
(\ie, \texttt{int x = 0}) was eliminated, and the miscompiled
executable just returns a memory cell with uninitialized data. This
bug is also a regression affecting the stable release of \clang 3.7
and all later versions.

\subsection{Toward Bounded Compiler Verification}\label{sec:ubdis}

As mentioned in Section~\ref{sec:intro}, our approach is general and
establishes the first step toward practical techniques for proving the
absence of compiler bugs for any programming language.
For C/C++ compilers, the SPE technique
itself does not guarantee that the generated programs are free of undefined behaviors.
Specifically, for the incorrect return value of the program described in Figure~\ref{fig:motivating-example-gcc-wrong-code}, our technique cannot 
determine directly whether it is a compiler miscompilation or a false alarm due to
possible undefined behavior. We rely on the heuristics discussed in Section~\ref{sec:expsetup}
and manual inspection to confirm the bug.
The test program was generated by SPE, which we believe can help prove the absence of                                                             
miscompilations in C/C++ compilers.   
Our SPE technique has indeed found several wrong code bugs in both GCC and Clang, but much fewer than 
crash bugs. This section briefly discusses 
practical considerations in finding wrong code bugs with skeletal
program enumeration.

The most significant challenge is to avoid enumerating programs with
undefined behaviors. In both program generation and mutation,
one can design different heuristics to avoid producing ``bad''
programs. For instance, when performing statement insertions in
\textsf{Athena}~\cite{LeSS15finding}, one can carefully
choose the candidate statements to avoid introducing
undefined behavior such as uninitialized variables or out-of-bound
array accesses. Moreover, a key contribution of
Csmith~\cite{YangCER11finding} is to ensure that its generated
programs are, most likely, free of undefined behavior. However, in
our SPE work, what heuristics to use is less obvious since we 
consider all variable usage patterns. On the other hand, SPE is
deterministic and exhaustive rather than opportunistic. As a result,
applying static analysis on each enumerated program would not be too
expensive. We leave as interesting future work to explore 
static analysis techniques or efficient enumeration schemes to avoid
undefined behaviors in the enumerated programs.

Besides avoiding undefined behaviors, it is also challenging to detect
undefined behaviors given a set of enumerated programs. This is
perhaps more general since the issue itself has been an interesting,
actively researched problem.  The reference interpreter in
CompCert~\cite{Leroy06formal}, for example, offer tremendous help in
detecting ``bad'' programs. However, since CompCert only works on a
subset of C, it may not handle many practically useful features such
as inline assembly, attributes and compiler-specific extensions.  It
also defines certain undefined behaviors, such as signed integer
overflows.  Tools such as Clang's undefined behavior sanitizers are
also useful, but incomplete. As a result, we resort to manual
inspection to rule out the remaining ``bad'' programs, which hinders
productivity. Reliable tools for detecting undefined behaviors would
be extremely helpful.


\section{Related Work}\label{sec:rw}

Csmith is the most popular random program generator for testing C
compilers~\cite{ChenGZWFER13taming,YangCER11finding}. Compared with
the testsuite used in our study, Csmith generates large and complex
programs. Csmith is a highly influential project. Over the years, it
has helped find a few hundred bugs in both GCC and Clang/LLVM. Based on
Csmith, the CLsmith work of Lidbury~\myetal\ focuses on testing OpenCL
compilers~\cite{LidburyLCD15many}. Orange3 is a random
program generator that tests arithmetic optimizations in C
compilers~\cite{NagaiHI14reinf}. CCG is another random C program
generator which finds crash bugs in early versions of GCC and
Clang~\cite{ccg}. 
Epiphron is a randomized technique to detect defects in compiler warning
diagnostics in GCC and Clang~\cite{SunLS16ICSE}.
For functional languages, there has also been
an extensive body of work on exhaustive or random test-case generators for
compiler testing~\cite{ClaessenH00quick,ClaessenDP15generating,FetscherCPHF15making,PalkaCRH11testing,DuregardJW12feat,RuncimanNL08smallcheck}. 
Boujarwah and Saleh conduct a thorough survey on 
generation techniques for compiler testing~\cite{BoujarwahS97compiler}.

A recent work of Le~\myetal\ proposes the idea of testing compilers
using the equivalence modulo inputs (EMI)~\cite{LeAS14compiler}
concept. Practical testing tools based on EMI mutate programs by
inserting and deleting statements in unexecuted branches. In
particular, \textsf{Orion} randomly deletes program statements in dead
regions~\cite{LeAS14compiler}. \textsf{Athena} adopts the Markov Chain
Monte Carlo (MCMC) method to guide both statement insertions and
deletions to obtain more interesting test
programs~\cite{LeSS15finding}. 
\textsf{Hermes} inserts code fragments to live regions~\cite{SunLS16finding}.
Moreover, \textsf{Proteus} applies the
EMI technique to test link-time optimizers~\cite{LeSS15rand}. The
frameworks based on EMI are quite efficient for compiler testing. They
have revealed many bugs in both GCC and Clang/LLVM. Most of them
are deep wrong code bugs. Besides testing C compilers, LangFuzz
mutates syntactically correct JavaScript programs using failing code
fragments~\cite{HollerHZ12fuzzing}. It has discovered many
vulnerabilities in the Mozilla JavaScript interpreter. Finally, the
well-known mutation testing technique mutates a program to evaluate
the quality of its testsuite~\cite{DeMilloLS78hints,Hamlet77testing}.

To guarantee the correctness of compilers, the two most notable
developments are, perhaps, translation
validation~\cite{Necula00tran,PnueliSS98tran} and verified
compilers~\cite{Leroy06formal}. Besides verification, compiler testing
is another important practical approach. For testing C compilers, all
of the program generation, program mutation and our SPE techniques realize the same idea of differential
testing~\cite{McKeeman98diff}. The three approaches complement each
other. Specifically, for program enumeration, we consider small test
programs. Our technique exhaustively exploits all variable
combinations. On the other hand, the other two approaches tend to
produce large and complex programs in a randomized fashion. The buggy
programs discovered using these techniques could be processed using
CompCert's reference interpreter to identify undefined
behaviors~\cite{Leroy06formal}. To file high-quality bug reports,
test programs should also be reduced first, using tools
like C-Reduce~\cite{RegehrCCEEY12test} and Berkeley
Delta~\cite{deltadan}.

Our work is also related to bounded-exhaustive testing, which concerns
the enumeration of all possible input structures up to a given
size~\cite{SullivanYCKJ04soft}. Two popular techniques are declarative
enumeration and imperative enumeration. In particular, declarative
approaches leverage any given invariant to search for valid
inputs~\cite{BoyapatiKM02korat,KhurshidM04testera,SenniF12genera,GaleottiRPF13taco},
and the imperative approaches directly construct the inputs based on
more prescriptive
specifications~\cite{VisserPK04test,RuncimanNL08smallcheck,DanielDGM07automated,KurajKJ15programming}. In
program synthesis, there have been studies on inductive functional
programming systems for exhaustively synthesizing small
programs~\cite{BriggsO08functional,Katayama08efficient,Katayama05systematic,Katayama12an}. The
essential enumeration techniques, categorized as analytical or
generate-and-test approaches, share similar conceptual ideas. As
mentioned in Section~\ref{sec:pediss}, existing enumeration techniques
are expensive and impractical for the combinatoral enumeration problem
that this work considers.

\section{Conclusion}\label{sec:ff}

This paper has introduced skeletal program enumeration (SPE) for
compiler testing and developed a practical combinatorial solution.
Our approach significantly reduces the number of enumerated programs.
For an empirical demonstration of its utility, we have
applied it to test two production C/C++ compilers, CompCert C comipler and two Scala compilers. Our results are extremely promising. For instance, in less than six
months, our approach has helped discover more than 200 bugs in GCC and Clang. More than 
half of our reported bugs have already been fixed, the majority are long latent,
and a significant fraction are classified as critical, release
blocking.

Our SPE strategy and techniques are general and may be applied
in other enumeration settings. This work also demonstrates the
practical potential of program enumeration, and opens up opportunities
toward bounded compiler verification.

\acks

We would like to thank our shepherd, Mayur Naik, and the anonymous PLDI
reviewers for valuable feedback on earlier drafts of this paper, which helped
improve its presentation. We thank Yang Chen and Davide Italiano for helpful comments on this work.
This research was supported in part by the United States National Science
Foundation (NSF) Grants 1319187, 1528133, and 1618158, and by a Google Faculty
Research Award. The information presented here does not necessarily reflect the
position or the policy of the Government and no official endorsement should be
inferred.

\ifapps
\appendix

\begin{figure*}[!t]
\centering

\subfigure[figbotcap][\gcc crash/performance bug \anonybugid{67619} (fixed)]{
  \usebox\myboxGccSampleTwo
\label{subfig:sample:gcc:67619}
}
\hfill
\subfigure[figbotcap][\gcc wrong code bug \anonybugid{70138} (fixed)]{
  \usebox\myboxGccSampleFour
\label{subfig:sample:gcc:70138}
}   
\hfill
\subfigure[figbotcap][G++ crash bug \anonybugid{71405} (fixed)]{
	\usebox\myboxGccSampleSeven
	\label{subfig:sample:71405}
}

\parbox[t]{\textwidth}{
    \parbox{.32\textwidth}{%
        \subfigure[Clang++ crash bug \anonybugid{28045} (fixed)]{
        	\usebox\myboxClangSampleEight
        	\label{subfig:sample:28045}
        } 
    }
    \hfill
    \parbox{.35\textwidth}{%
        \subfigure[CompCert crash bug \anonybugid{125} (fixed)]{
        	\usebox\myboxCcompSampleOne
        	\label{subfig:sample:125}
        }
        
        \subfigure[Dotty crash bug \anonybugid{1637} (fixed)]{
        	\usebox\myboxDottySampleTwo
        	\label{subfig:sample:1637}
        }
    }   
    \hfill
    \parbox{.29\textwidth}{%
        \subfigure[CompCert crash bug \anonybugid{121} (fixed)]{
        	\usebox\myboxCcompSampleTwo
        	\label{subfig:sample:121}
        }
        \vspace*{0.36cm}
        
        \subfigure[Scala crash bug \anonybugid{10015} (open)]{
        	\usebox\myboxScalaSampleOne
        	\label{subfig:sample:10015}
        }
    } 
}

\caption{Additional sample bugs.} \label{fig:appbug}
\end{figure*}

\section{Additional Sample Bugs} \label{app:additional}

We briefly discuss eight additional sample bugs found by SPE to show
its generality and the diverse bugs that it can detect.

\paragraph{Figure~\ref{subfig:sample:gcc:67619}.}
This bug is long latent and intriguing as it causes different 
symptoms for multiple \gcc versions. 
It affects optimization levels -O1 and above.
When compiling it, 
\gcc 4.6 and 4.7 hang, whereas 
4.8 to trunk crash. 
\gcc incorrectly computes the address for exception handling, which later
causes an assertion violation in the middle end.

\paragraph{Figure~\ref{subfig:sample:gcc:70138}.}
This test program is miscompiled by the \gcc trunk (6.0 revision
234026) at -O3.  It is derived by enumerating a test case in
\gcc's testsuite.  The expression \codeIn{c=c+u[a+1335*a]} on
line~\ref{line:gcc:70318:replacement} is obtained by
replacing \codeIn{b} in the original expression
\codeIn{c=c+u[a+1335*b]} with \texttt{a}. 
Then this replacement triggers a regression in the loop 
vectorizer pass. 

\paragraph{Figure~\ref{subfig:sample:71405}.}
This test program crashes the trunk (7.0 revision 237059) of \gcc at
-Os and above.  The code overrides the placement new operator of C++
on line~\ref{line:gcc:71405:replacement}. A replacement new operator
creates an object in a given memory region.  In the main function,
this overridden new operator is called to create an object of
type \codeIn{C} at the address of the local variable \codeIn{a} (\ie,
`\codeIn{new (\&a) C}').  However, because \codeIn{C} and \codeIn{a}
have different types, \gcc translates the code into an ill-formed
intermediate representation (\ie, GIMPLE code), which does not pass
the GIMPLE verification pass.

\paragraph{Figure~\ref{subfig:sample:28045}.}
This program triggers a bug in the name mangling module of \clang for
the Itanium C++ ABI.  On line~\ref{line:clang:28045:test}, the
template function \codeIn{test} takes as input two parameters of the
types: a generic type \codeIn{T} and the underlying type
of \codeIn{T}.  When \clang was trying to mangle the function name of
the call on line 11, the bug (\ie, the type
trait \codeIn{\_\_underlying\_type} was improperly handled) led the
compilation to unreachable code, thus failing an assertion.

\paragraph{Figure~\ref{subfig:sample:125}.}
This test program triggers a crashing bug in CompCert's frontend.
Before the initialization, the parser does not check whether the type is
incomplete, which triggers an assertion failure in CompCert.

\paragraph{Figure~\ref{subfig:sample:1637}.}
This test program crashes the Dotty compiler --- a next generation
compiler for Scala. It triggers an assertion in the Dotty typer. The
bug has been fixed and marked as high priority in Dotty's GitHub
repository. As of March 2017, there are five high-priority bugs 
in the Dotty code repository, and SPE discovered four.

\paragraph{Figure~\ref{subfig:sample:121}.}
This test program crashes CompCert. Function \texttt{foo}'s parameter has a
structure type \codeIn{A}, whose definition is unavailable in this
translation unit. CompCert did not reject the program early, thus
leading to an ``Unbound struct A'' assertion failure in the subsequent
compilation of the program.

\paragraph{Figure~\ref{subfig:sample:10015}.}
This test program crashes the 2.12 stable release of the Scala
compiler. Specifically, it triggers an assertion failure
in Scala's type checker. The test program is enumerated from the
regression test-suite in the Scala release.

\fi

\bibliographystyle{abbrvnat}
\balance 
\bibliography{bib/oopsla2016}

\end{document}